\documentclass[aps,pre,superscriptaddress,showpacs,twocolumn]{revtex4}
\usepackage{dcolumn}
\usepackage{graphicx}
\usepackage{amssymb}
\usepackage{amsmath}
\usepackage{enumerate}

\begin{document}

\title{Interpenetration as a Mechanism for\\Liquid-Liquid Phase Transitions}

\date{\today}

\author{Chia Wei Hsu}
\affiliation{Department of Physics, Wesleyan University, Middletown, Connecticut 06459, USA}

\author{Francis W. Starr}
\affiliation{Department of Physics, Wesleyan University, Middletown, Connecticut 06459, USA}

\begin{abstract}

  We study simple lattice systems to demonstrate the influence of
  interpenetrating bond networks on phase behavior. We promote
  interpenetration by using a Hamiltonian with a weakly repulsive
  interaction with nearest neighbors and an attractive interaction with
  second-nearest neighbors. In this way, bond networks will form between
  second-nearest neighbors, allowing for two (locally) distinct networks
  to form. We obtain the phase behavior from analytic solution in the
  mean-field approximation and exact solution on the Bethe lattice. We
  compare these results with exact numerical results for the phase
  behavior from grand canonical Monte Carlo simulations on square,
  cubic, and tetrahedral lattices. All results show that these simple
  systems exhibit rich phase diagrams with two fluid-fluid critical
  points and three thermodynamically distinct phases. We also consider
  including third-nearest-neighbor interactions, which give rise to a
  phase diagram with four critical points and five thermodynamically
  distinct phases. Thus the interpenetration mechanism provides a simple
  route to generate multiple liquid phases in single-component systems,
  such as hypothesized in water and observed in several model and
  experimental systems. Additionally, interpenetration of many such
  networks appears plausible in a recently considered material made from
  nanoparticles functionalized by single strands of DNA.

\end{abstract}

\date{Submitted: November 18, 2008}


\maketitle

\section{Introduction}

In the last 15 years, liquid-liquid phase transitions in one-component
systems has been an area of vigorous
research~\cite{pgam,ms98review,ds-review}. Liquid-liquid phase
transitions have been found experimentally in
phosphorus~\cite{phosphorus}, and are also suspected for many
tetrahedrally coordinated fluids including water~\cite{pses92, ms98,
  sslss, debenedetti-review, brovchenko, pss}, carbon~\cite{carbon},
silica~\cite{silica1,silica2}, and silicon~\cite{silicon}. A variety of
different approaches have been used to understand the emergence of a
second liquid state, most of which rely on a competition between
non-directional van der Waals interactions and directional bonding
interactions that favor more open
states~\cite{sss,psgsa,sing-free,roberts,fms}. Some success has also
been found for models with purely symmetric
interactions~\cite{sbfms,bs,fmsbs,xu06}. Recently, a model for
nanoparticles functionalized by single strands of DNA~\cite{ss,lss,lst}
whose sequence promotes bonding between the nanoparticle units showed
that polyamorphic behavior can also arise due solely to directional
interactions that result in open networks which interpenetrate, giving
up to three critical points and four amorphous phases~\cite{hlss}. The
question that remains is, can network interpenetration be a general
mechanism to liquid-liquid phase transitions?

Lattice models have long served to provide analytic insight into complex
physical problems~\cite{baxterBook,HESbook,WuReview}. The most famous
example is the Ising model, which helped to understand the origin of
spontaneous magnetization in magnetic systems. The Ising model can also
be recast as a lattice gas model, which equivalently provides an
understanding of condensation of the liquid from the gas. The Ising model
(or lattice gas) exhibits such complexity while only using very simple
first neighbor interactions. Making these interactions or lattice
occupancy state more complex -- for example, the Potts model, the
spherical model, etc -- can yield even richer
behavior~\cite{baxterBook}.

Indeed, lattice model approaches based on a lattice gas with additional
orientation dependent bonding
interactions~\cite{sss,sing-free,roberts,fms} have reproduced phase
behavior with two critical points. Motivated by the finding that
interpenetration might serve as a mechanism for generating liquid-liquid
transitions, we explore lattice models that incorporate interpenetration
to see how phase behavior is affected. In our models we promote
interpenetration of networks by using an attractive interaction with
second-nearest neighbors (2NN) and a weakly repulsive interaction with
nearest neighbors (NN). The properties of Ising and lattice-gas models
with NN and 2NN interactions were first examined on a two dimensional
lattice, primarily for the case of purely antiferromagnetic interactions
({\it i.e.} both NN and 2NN
repulsions)~\cite{landau,bl,bl85,mas,mkt,mk,avs}.  Our case (repulsive NN
and attractive 2NN interactions) has been extensively studied for the 2D
triangular
lattice~\cite{triangular-landau,triangular-qb,triangular-mfk}, where the
results have been applied to gas adsorption on surfaces.  Considerably
less attention has been given to 3D lattices; most work has focused on
the zero-field Ising formulation of the model on the cubic
lattice~\cite{anjos07,db03,cgp97}, where both the phase behavior and
critical properties have been examined.  The phase behavior for the
lattice-gas formulation has also been examined on the hexagonal
close-packed lattice with application to binary metal
alloys~\cite{so00}.  Here we examine the lattice-gas formulation of the
model on 3D lattices including the cubic lattice and the diamond
lattice.  In the Ising formulation, this is equivalent to including a
non-zero external field.  For the completeness of our work, we also
consider the behavior on the 2D square lattice.  For all three lattices
we carry out both the mean field solution and exact numerical Monte
Carlo (MC) solution. For the symmetric lattice model, we also provide an
exact solution on the Bethe lattice.

Our findings indicate that on all lattices there are two critical points
and three distinct phases: (i) unassociated molecules of gas, (ii)
liquid I, a single network of alternatingly filled sites, and (iii)
liquid II, a double interpenetrating network with all sites occupied.
For all lattice, the two critical temperatures are the same due to the
symmetry between occupied and empty sites.  To connect more closely with
systems of experimental interest, we also consider a slightly more
complicated model that includes a three-body term that accounts for
increased repulsion in crowded states.  This model breaks the symmetry
between occupied sites. For this model, we find that the high density
critical point occurs at a temperature lower than the low density
critical point, as observed in most polyamorphic systems.

The Bethe lattice has a distinct geometry from other lattices we study,
and it exhibits somewhat different phase behavior. While there are still
two critical points and three distinct phases, there is at least one
case in which the coexistence lines merge for temperatures slightly
lower than $T_c$ and separate again at even lower temperatures. In
addition, a density anomaly occurs, as does in water~\cite{debenedetti-review}.

Since it has been argued that the observation of even more amorphous
phases can result from very open network structures~\cite{hlss}, we also
consider a modification of the previous model. Specifically, we consider
a model including third-nearest-neighbor (3NN) attraction, and weak
first and second neighbor repulsion~\cite{bl85}. Results on the square lattice show
that there can be up to 4 critical points for this system.

The paper is organized as follows: Sec.~\ref{model} describes the three
models: (i) the symmetric model, (ii) the asymmetric model, and (iii) the 3NN
interaction model. Sec.~\ref{method} gives the derivation of the
mean-field approximation and Bethe lattice solution, and describes the
methods used in the MC simulations. Sec.~\ref{result} presents the
results and discussion. We conclude briefly in Sec.~\ref{conclusion}.


\section{Models}
\label{model}

\subsection{Symmetric Lattice Model}
\label{model-sym}

The first and simplest model we consider is a second neighbor lattice gas with Hamiltonian
\begin{equation}
\label{eq:sym2NN}
\mathcal{H}=-\epsilon_1\sum_{NN}n_in_j-\epsilon_2\sum_{2NN}n_in_j\quad,
\end{equation}
where $\sum_{NN}$ indicates a sum over all NN pairs and $\sum_{2NN}$ indicates a sum over all 2NN pairs. At each site $i$, the occupancy $n_i$ is $1$ when the site is occupied and $0$ when the site is unoccupied. The volume $V$ of this system is the total number of sites taking the volume of a single site $v=1$. We define the ratio of interaction strength $R=\epsilon_1/\epsilon_2$.

To promote interpenetration, we choose $\epsilon_1<0$ and $\epsilon_2>0$. In this way, bond networks will form between second-nearest neighbors, allowing for two (locally) distinct networks to form.
In nature, this phenomenon occurs in ices VI, VII and VIII, where two interpenetrating structures form. Multiple interpenetrating networks also occur in model of DNA-functionalized nanoparticles, and gives rise to polyamorphic phase behavior~\cite{hlss}. The aim of our lattice model is to retain this interpenetration feature while eliminating other complexities. We expect this simple lattice model to exhibit multiple critical points and to serve as a demonstration of the influence of interpenetrating bond networks on phase behavior.

This lattice model (like the Ising model) has an intrinsic symmetry
between occupied states and unoccupied states; thus we call this the
symmetric model. 

\subsection{Asymmetric Lattice Model}
\label{model-asym}

The intrinsic symmetry between occupied and unoccupied states is usually not found in nature. To break the particle/hole symmetry, we introduce a three-body term in the NN interaction. Such three-body interactions for the antiferromagnetic case have been studied elsewhere~\cite{3body82,3body83}. This term accounts for the increased repulsion as crowding occurs. One can think of this three-body interaction as scaling the NN interaction strength by the ratio of number of occupied sites around the pair to the number of sites around the pair. This leads to the Hamiltonian
\begin{equation}
\label{eq:asym}
\mathcal{H}=-\epsilon_1\sum_{NN}n_in_j\sum_{<ijk>}\frac{n_k}{c}-\epsilon_2\sum_{2NN}n_in_j,
\end{equation}
where $\sum_{<ijk>}$ sums over all sites $k$ that are first neighbor of either site $i$ or site $j$; $c$ is the total number of such sites for any pair $(i,j)$. On the 2D square lattice and 3D diamond lattice, $c=6$; on a cubic lattice $c=10$.

\subsection{Third-neighbor Interaction Model}
\label{model-3NN}

We shall see that the lattice model with only first and second neighbor interactions cannot account for more than two critical points. Hence, as an extension, we consider a model including up to third neighbor interactions. 
\begin{equation}
\label{eq:sym3NN}
\mathcal{H}=-\epsilon_1\sum_{NN}n_in_j-\epsilon_2\sum_{2NN}n_in_j-\epsilon_3\sum_{3NN}n_in_j
\end{equation}
As in previous models, we choose $\epsilon_1<0$, $\epsilon_2<0$, and $\epsilon_3>0$ to promote interpenetration. We define two ratios of interaction strength: $R_1=\epsilon_1/\epsilon_3$ and $R_2=\epsilon_2/\epsilon_3$. This model is more representative of systems in which the bonding is very large in comparison to the core repulsion. This third neighbor model for the purely antiferromagnetic case ($\epsilon_1<0$, $\epsilon_2<0$, $\epsilon_3<0$) has also previously been examined~\cite{bl85}.


\section{Methods}
\label{method}

\subsection{Mean Field Approximation}
\label{method-MF}

\subsubsection{Symmetric Case}
\label{method-MF-sym}

\begin{figure}
\begin{center}
\includegraphics[width=3in]{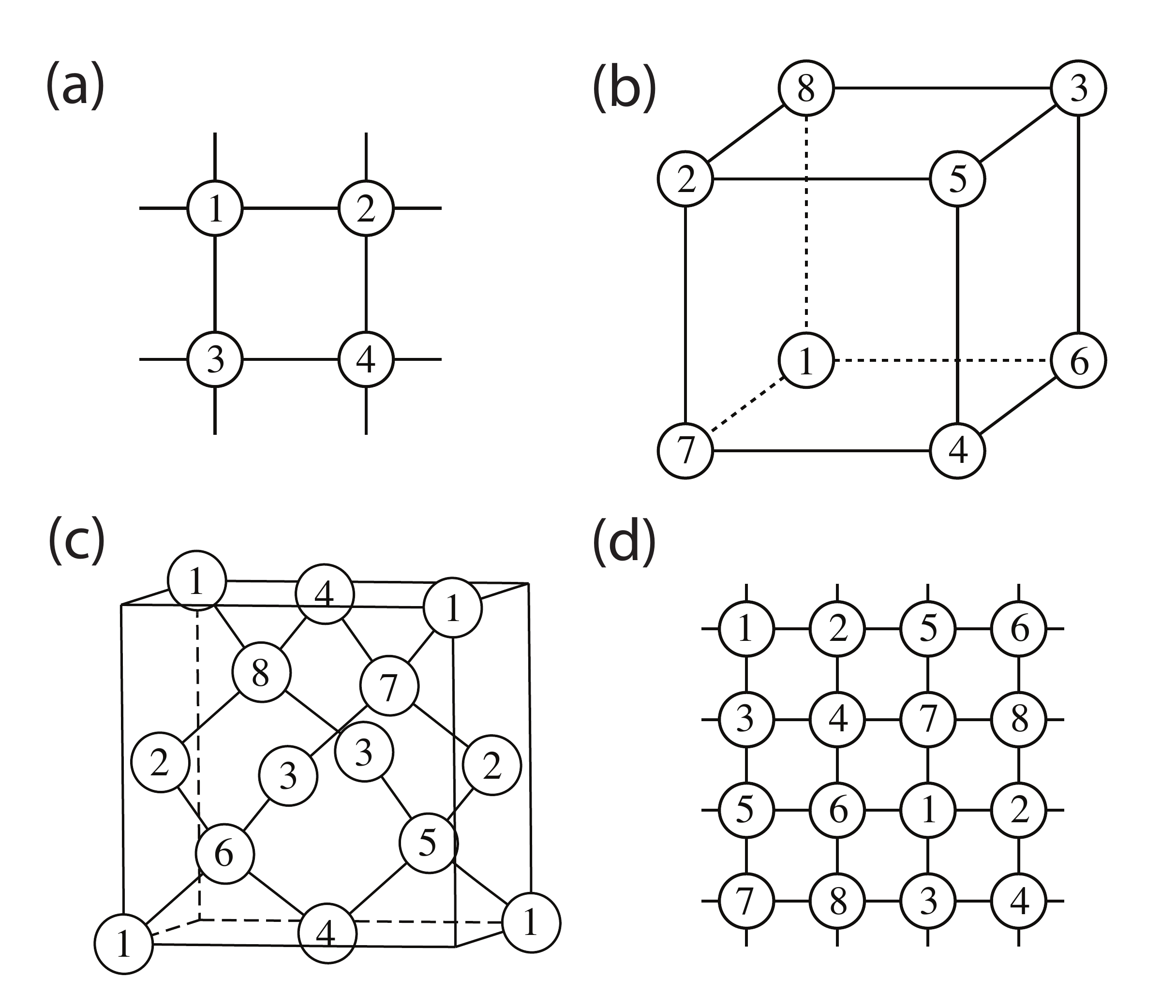}
\end{center}
\caption{Illustration of sub-lattice division for the MF approximation. (a) 2D square lattice. (b) cubic lattice. (c) diamond lattice. (d) 2d square lattice with 3NN interaction. For (a)-(c) the division is made such that there is no NN or 2NN interaction between sites on the same sub-lattice. For (d) the division is made such that there is no NN, 2NN, or 3NN interaction between sites on the same sub-lattice. }
\label{fig:division}
\end{figure}

To carry out the mean-field solution, we follow the procedure described by Binder and Landau~\cite{bl}. Specifically, we first divide the lattice into sub-lattices such that sites 

\begin{table}[h]
\caption{List of relevant values for the MF approximation. $f(\alpha,\beta)$ and $g(\alpha,\beta,\gamma)$ are used for the asymmetric model. }
\label{table:sublattice}
\begin{center}
\begin{tabular}{p{1.0in}|p{0.2in}p{0.2in}p{0.2in}p{0.2in}|p{0.6in}p{0.6in}}
\hline\hline
lattice & $\nu$ & $\gamma_1$ & $\gamma_2$ & $\gamma_3$ & $f(\alpha,\beta)$ & $g(\alpha,\beta,\gamma)$ \\
\hline
square & 4 & 2 & 4 & & 1 & 2 \\
cubic & 8 & 2 & 4 & & 1 & 0 or 2\\
diamond & 8 & 1 & 4 & & 0 & 1\\
square (3NN) & 8 & 1 & 2 & 4 & &\\
\hline\hline
\end{tabular}
\end{center}
\end{table}
\noindent on the same sub-lattice do not interact with each other. The way we make such division for the three lattices is illustrated in Fig.~\ref{fig:division}. We denote the number of such sub-lattices $\nu$. We define the interaction energy between a site on sub-lattice $\alpha$ and a site on sub-lattice $\beta$ as $\epsilon_{\alpha\beta}$. We denote the number of such $(\alpha,\beta)$ pairs for a site on either sub-lattice as $\gamma_{\alpha\beta}$. If the $\alpha$ and $\beta$ sub-lattices are nearest neighbors, $\epsilon_{\alpha\beta}=\epsilon_1$ and $\gamma_{\alpha\beta}=\gamma_1$. If $\alpha$ and $\beta$ sub-lattices are second-nearest neighbors, $\epsilon_{\alpha\beta}=\epsilon_2$ and $\gamma_{\alpha\beta}=\gamma_2$. Otherwise there is no interaction so $\epsilon_{\alpha\beta}=0$ and $\gamma_{\alpha\beta}=0$. The values for $\nu$, $\gamma_1$, and $\gamma_2$ for the three lattices we study are listed in Table~\ref{table:sublattice}.

In mean-field approximation, the occupancy of a neighboring site is approximated as the mean density of the sub-lattice to which it belongs. There are $V/\nu$ such sites on each sub-lattice, thus the mean-field Hamiltonian is
\begin{equation}
\label{eq:MF-sym-H}
\mathcal{H}^{MF}=-\frac{V}{\nu}\sum_{\alpha,\beta}\gamma_{\alpha\beta}\epsilon_{\alpha\beta}\rho_\alpha\rho_\beta=-\frac{\nu}{V}\sum_{\alpha,\beta}\gamma_{\alpha\beta}\epsilon_{\alpha\beta}N_{\alpha}N_\beta,
\end{equation}
where $\rho_\alpha=(N_\alpha\nu)/V$, and $N_\alpha=\sum_{i \in \alpha}n_i$ is the number of occupied sites on sub-lattice $\alpha$. To avoid double counting, the sum $\sum_{\alpha,\beta}$ is specifically $\sum_{\alpha=1}^{\nu}\sum_{\beta=\alpha+1}^{\nu}$. The mean-field Hamiltonian depends on the configuration via $N_1,N_2,...N_\nu$. We label this set $\{N_\alpha\}$ for brevity.

We fix $\{N_\alpha\}$, $V$, $T$, and evaluate the canonical partition function
\begin{widetext}
\begin{equation}
\label{eq:MF-sym-Z}
\mathcal{Z}\left(\{N_{\alpha}\},V,T\right)=\left[\prod_{\alpha}\binom{V/\nu}{N_\alpha}\right]
\exp\left(\frac{\nu}{V}\sum_{\alpha,\beta}\gamma_{\alpha\beta}\epsilon_{\alpha\beta}N_{\alpha}N_\beta/kT\right).
\end{equation}
\end{widetext}
Using the thermodynamic relations
\begin{equation}
\label{eq:diff}
\mu_\alpha=-kT\left(\frac{\partial{\ln}\mathcal{Z}}{{\partial}N_\alpha}\right)_{\{N_{\beta\neq\alpha}\},V,T}\qquad P=kT\left(\frac{\partial{\ln}\mathcal{Z}}{{\partial}V}\right)_{\{N_\alpha\},T}
\end{equation}
and Stirling's approximation, we find
\begin{equation}
\label{eq:MF-sym-mu}
\mu_\alpha=kT\ln\frac{\rho_\alpha}{1-\rho_\alpha}-\sum_{\beta}\gamma_{\alpha\beta}\epsilon_{\alpha\beta}\rho_\beta,
\end{equation}
\begin{equation}
\label{eq:MF-sym-P}
P=-\frac{kT}{\nu}\sum_{\alpha}\ln\left(1-\rho_\alpha\right)-\frac{1}{\nu}\sum_{\alpha,\beta}\gamma_{\alpha\beta}\epsilon_{\alpha\beta}\rho_\alpha\rho_\beta.
\end{equation}

Ref.~\cite{bl} derives this result using the grand canonical partition function. Using the canonical approach will be particularly useful when we introduce three-body terms.

Rearrangement of Eq.~(\ref{eq:MF-sym-mu}) leads to
\begin{widetext}
\begin{equation}
\label{eq:MF-sym-rho}
\rho_\alpha={\displaystyle \exp\left[\left(\sum_{\beta}\gamma_{\alpha\beta}\epsilon_{\alpha\beta}\rho_\beta-\mu\right)/kT\right]}\left/\left\{ 1+\displaystyle \exp\left[\left(\sum_{\beta}\gamma_{\alpha\beta}\epsilon_{\alpha\beta}\rho_\beta-\mu\right)/kT\right]\right\} \right.,
\end{equation}
where $\alpha$ goes from 1 to $\nu$. Eq.~(\ref{eq:MF-sym-rho}) represents an system of $\nu$ implicit equations for sub-lattice densities $\{\rho_\alpha\}$. Given $\mu$ and $T$, we obtain the set $\{\rho_\alpha\}$ by  solving this system of equations numerically using the iterative scheme described in Ref.~\cite{bl}. Specifically, we use

\begin{equation}
\label{eq:MF-sym-iterate}
\rho_\alpha^{(n)}=(\cos^2\phi)\rho_\alpha^{(n-1)}+
(\sin^2\phi){\displaystyle \exp\left[\left(\sum_{\beta}\gamma_{\alpha\beta}\epsilon_{\alpha\beta}\rho_\beta-\mu_\alpha\right)/kT\right]}\left/\left\{1+\displaystyle \exp\left[\left(\sum_{\beta}\gamma_{\alpha\beta}\epsilon_{\alpha\beta}\rho_\beta-\mu_\alpha\right)/kT\right]\right\}\right.,
\end{equation}
\end{widetext}
where $\phi$ is an arbitrary tuning parameter. When $\phi=0$, the series stays at $\rho_{\alpha}^{(0)}$. When $\phi=\pi/2$, $\rho_\alpha$ is a direct iteration of Eq.~(\ref{eq:MF-sym-rho}). One needs to choose appropriate $\phi$ so that the series converges. As discussed in Ref.~\cite{bl}, this iterative scheme prevents us from finding unstable solutions which are not physical. We choose the initial densities $\{\rho_\alpha\}$ and iterate using Eq.~(\ref{eq:MF-sym-iterate}) until $|\rho_\alpha^{(n)}-\rho_\alpha^{(n-1)}|<\delta$ for all $\alpha$. In our calculations the tolerance $\delta=10^{-9}$, and most of the time we choose $\phi=\pi/4$.

Eq.~(\ref{eq:MF-sym-rho}) can have multiple solutions. To find all possible stable or metastable solutions, we repeat the iteration procedure with different sets of initial densities. In the cases when multiple solutions are found, we calculate the pressure using Eq.~(\ref{eq:MF-sym-P}), and take the solution with the highest pressure ({\it i.e.} lowest grand potential free energy $\Omega$), which is the most thermodynamically stable state. We can distinguish first order transitions by a discontinuity in density ({\it i.e.} $\frac{\partial P}{\partial \mu}$), and second order transitions by a discontinuity in the slope of density ({\it i.e.} $\frac{\partial^2 P}{\partial \mu^2}$).

\subsubsection{Asymmetric Case}
\label{method-MF-asym}

For the asymmetric model, the NN and 2NN interactions have a different form from each other, and we must write out the NN and 2NN terms separately, unlike the symmetric case. The mean-field Hamiltonian is
\begin{equation}
\mathcal{H}^{MF}=-\frac{V}{\nu}\left[\frac{\gamma_1\epsilon_1}{c}\sum_{NN}\rho_\alpha\rho_\beta\sum_{<\alpha\beta\delta>}\rho_\delta+\gamma_2\epsilon_2\sum_{2NN}\rho_\alpha\rho_\beta\right],
\end{equation}
where $\sum_{NN}$ sums over all distinct pairs of sub-lattices $(\alpha,\beta)$ that are first neighbors, $\sum_{2NN}$ sums over all distinct pairs of sub-lattices $(\alpha,\beta)$ that are second neighbors, and $\sum_{<\alpha\beta\delta>}$ sums over all $c$ sub-lattices $\delta$ that are a first neighbor of sub-lattice $\alpha$ or $\beta$. It is possible that a particular sub-lattice may appear twice as a neighbor to lattice $\alpha$ and $\beta$, so some sub-lattices must be counted twice in this summation. 

Now rewrite the Hamiltonian as a function of $\{N_\alpha\}$,
\begin{widetext}
\begin{equation}
\mathcal{H}^{MF}=-{\left(\frac{\nu}{V}\right)}^2\frac{\gamma_1\epsilon_1}{c}\sum_{NN}N_{\alpha}N_\beta\sum_{<\alpha\beta\delta>}N_\delta-\left(\frac{\nu}{V}\right)\gamma_2\epsilon_2\sum_{2NN}N_{\alpha}N_\beta,
\end{equation}
and write out the canonical partition function, fixing  $\{N_\alpha\}$, $V$, and $T$,
\begin{equation}
\label{eq:MF-asym-Z}
\mathcal{Z}\left(\{N_{\alpha}\},V,T\right)=\left[\prod_{\alpha}\binom{V/\nu}{N_\alpha}\right]\exp\left\{\left[{\left(\frac{\nu}{V}\right)}^2\frac{\gamma_1\epsilon_1}{c}\sum_{NN}N_{\alpha}N_\beta\sum_{<\alpha\beta\delta>}N_\delta+\left(\frac{\nu}{V}\right)\gamma_2\epsilon_2\sum_{2NN}N_{\alpha}N_\beta\right]/kT\right\}.
\end{equation}

To calculate $\mu$, we take partial derivative of the summation $\sum_{NN}$ with respect to $N_\alpha$. When doing so, we need to consider that $N_\alpha$ can appear in the NN sum, as well as the three-body sum, so there will be 3 terms associated with the derivative of the near neighbor interactions. Accordingly, we obtain
\begin{equation}
\label{eq:MF-asym-mu}
\mu_\alpha=kT\ln\frac{\rho_\alpha}{1-\rho_\alpha}-\frac{\gamma_1\epsilon_1}{c}\left[
\sum_{\alpha,NN}\rho_\beta\left(\sum_{<\alpha\beta\delta>}\rho_\delta+\rho_\alpha f(\alpha,\beta)\right)+\sum_{\substack{NN \\ \beta\neq\alpha\,,\delta\neq\alpha}}\rho_\beta\rho_\delta g(\alpha,\beta, \delta)
\right]-\gamma_2\epsilon_2\sum_{\alpha,2NN}\rho_\beta,
\end{equation}
\end{widetext}
where $f(\alpha,\beta)=\frac{\partial \sum_{<\alpha\beta\delta>}\rho_\delta}{\partial \rho_\alpha}$, and $g(\alpha,\beta,\delta)=\frac{\partial \sum_{<\beta\delta\sigma>}\rho_\sigma}{\partial \rho_\alpha}$. The summation $\sum_{\alpha,NN}$ sums over all $\beta$ that are first neighbors of $\alpha$, $\sum_{\substack{NN \\ \beta\neq\alpha\,,\delta\neq\alpha}}$ sums over all pairs $(\beta, \delta)$ where $\beta$ and $\delta$ are first neighbors of each other, but $\beta\neq\alpha$ and $\delta\neq\alpha$, and $\sum_{\alpha,2NN}$ sums over all $\beta$ that are second neighbor of $\alpha$.

$f(\alpha,\beta)$ quantifies the repulsion on $\alpha$ due to 3-body terms containing only $\alpha$ and $\beta$, and $g(\alpha,\beta,\delta)$ quantifies the repulsion on $\alpha$ due to 3-body terms containing $\alpha$, $\beta$, and $\delta$. For the 2D square lattice, $f(\alpha,\beta)=1$ and $g(\alpha,\beta,\delta)=2$. For the cubic lattice, $f(\alpha,\beta)=1$, and $g(\alpha,\beta,\delta)$ can be 0 or 2 depending on weather $\alpha$ is a neighbor to the pair $(\beta, \delta)$. For the diamond lattice, $f(\alpha,\beta)=0$ and $g(\alpha,\beta,\delta)=1$. These values are also listed in Table~\ref{table:sublattice}.

To calculate $P$, we take partial derivative of $\ln\mathcal{Z}$ with respect to $V$. Here we get an extra factor of 2 for the NN interaction term because of the power $(\nu/V)^2$, which arises from its three-body nature. We obtain
\begin{widetext}
\begin{equation}
\label{eq:MF-asym-P}
P=-\frac{kT}{\nu}\sum_{\alpha}\ln\left(1-\rho_\alpha\right)-\frac{2\gamma_1\epsilon_1}{c\nu}\sum_{NN}\rho_\alpha\rho_\beta\sum_{<\alpha\beta\delta>}\rho_\delta-\frac{\gamma_2\epsilon_2}{\nu}\sum_{2NN}\rho_\alpha\rho_\beta
\end{equation}
\end{widetext}

Similar to the symmetric case, Eq.~(\ref{eq:MF-asym-mu}) is a system of $\nu$ implicit equations for $\{\rho_\alpha\}$. We rearrange it into the form of Eq.~(\ref{eq:MF-sym-rho}) to carry out the iteration scheme of Eq.~\eqref{eq:MF-sym-iterate}. We repeat this iteration procedure with different sets of initial densities, and take the solution with the highest pressure (lowest $\Omega$).

\subsection{Solution on Bethe Lattice}
\label{method-Bethe}

While exact solution on most lattices is either difficult or impossible, the recursive nature of the Bethe lattice (which we define explicitly below) sometimes makes solution possible. The Bethe lattice usually offers a better approximation to a regular lattice with the same coordination than the conventional mean-field approximation, since it preserves correlations between local sites. 

\begin{figure}
\begin{center}
\includegraphics[width=2in]{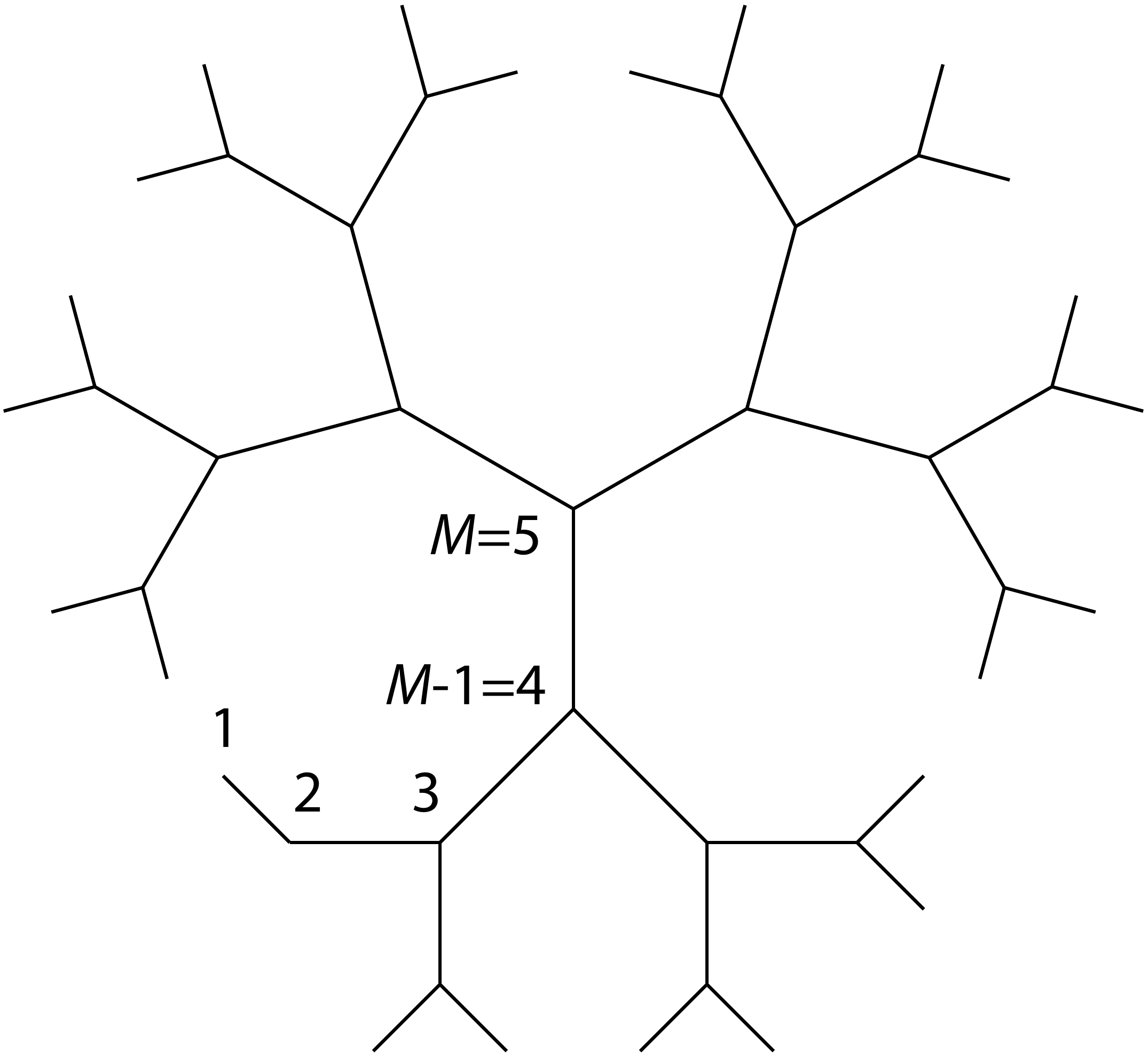}
\end{center}
\caption{Illustration of a Cayley tree with $\gamma=3$, $M=5$. Generation number is labeled on the graph.} 
\label{fig:bethe}
\end{figure}

A Cayley tree, as illustrated in Fig.~\ref{fig:bethe}, is constructed as follows: 
\begin{enumerate}[(i)]
\item We start from a center site and connect $\gamma$ sites to it. We say the center is generation $M$, and the $\gamma$ nearest-neighbor sites are generation $M-1$.
\item We connect $\gamma-1$ sites to each site in generation $M-1$. Repeat this procedure until we reach generation $1$.
\end{enumerate}

For most lattices, the ratio of surface sites to interior sites vanishes in the thermodynamic limit $M\to\infty$. However, on a Cayley tree, this ratio does not vanish~\cite{baxterBook}. The standard way to overcome this problem is to consider only sites that are infinitely far from the boundary in the thermodynamic limit. These central sites will not be affected by the surface, and they constitute the Bethe lattice~\cite{baxterBook}.

We now describe the solution to the symmetric lattice model on the Bethe lattice. We use $n_M$ to denote the occupancy for the center site and $\{n_{m}\}$ to denote the set of occupancy for sites in generation $m$. We use $\mathfrak{Z}_M$ to denote the grand partition function of the whole tree $\mathcal{T}_M$, and use $\mathfrak{z}_m$ to denote the partial grand partition function of a branch $\mathcal{B}_m$ whose top site is in generation $m$. When evaluating $\mathfrak{z}_m$, the occupancies $n_m$ and $n_{m-1}$ appear explicitly, so $\mathfrak{z}_m$ is a function of $n_m$ and $n_{m-1}$. Specifically,
\begin{widetext}
\begin{equation}
\mathfrak{z}_m(n_m, n_{m-1})=\sum_{\{n_i\}}\exp\left[\beta\left(\mu\sum_{i\in\mathcal{B}_m}n_i+{\epsilon_1}\sum_{NN\in\mathcal{B}_m}n_in_j+{\epsilon_2}\sum_{2NN\in\mathcal{B}_m}n_in_j\right)\right],
\end{equation}
\end{widetext}
where $\beta=1/kT$, not to be confused with the lattice enumerating variable in Sec.~\ref{method-MF}. $\sum_{\{n_i\}}$ sums over all possible configurations in branch $\mathcal{B}_m$, fixing $n_m$ and $n_{m-1}$. $\sum_{i\in\mathcal{B}_m}$ sums over all sites in $\mathcal{B}_m$ except for the top site in generation $m$.
$\mathfrak{z}_m$ does not account for 2NN interaction between different branches of the same generation $m$.

\begin{figure*}
\begin{center}
\includegraphics[width=7in]{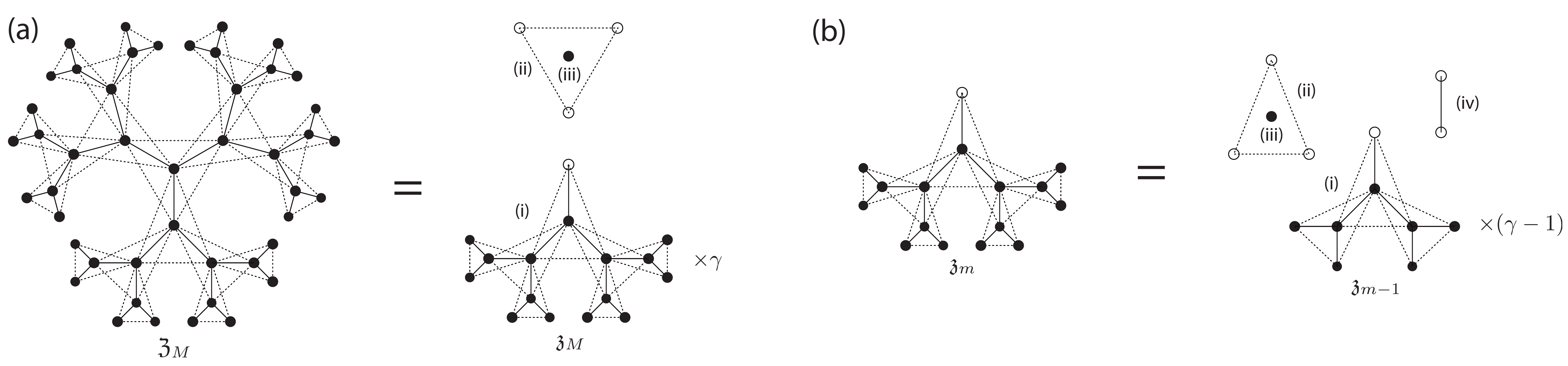}
\end{center}
\caption{Graphical representation of the separation of the Bethe lattice grand canonical partition function into different branches. We represent NN interaction with solid lines, 2NN interaction with dotted lines, and indicate the contribution from the chemical potential with filled circles. Empty circles represent sites where the chemical potential contribution is not counted. The left and right panels of (a) corresponds to the left and right hand sides of Eq.~(\ref{eq:ZM}). It shows how the total grand partition function $\mathfrak{Z}_M$ can be divided into (i) $\gamma$ partial grand partition functions $\mathfrak{z}_M$, (ii) 2NN interaction between the branches (represented as a triangle), and (iii) occupation of the center site (represented as a filled circle in the triangle). The left and right panels of (b) corresponds to the left and right hand sides of Eq.~(\ref{eq:zm0}). It shows how the partial grand partition function $\mathfrak{z}_m$ can be dived into (i) $\gamma-1$ partial grand partition functions of shorter branch $\mathfrak{z}_{m-1}$, (ii) 2NN interaction between generation $m$ and $m-2$ (represented as a triangle), (iii) occupation of the site in generation $m-1$ (represented as a filled circle in the triangle), and (iv) NN interaction between the site in generation $m$ and the site in generation $m-1$ (represented as a solid line).  }
\label{fig:separate}
\end{figure*}

The whole tree $\mathcal{T}_M$ is composed of $\gamma$ branches $\mathcal{B}_M$, so the total grand partition function $\mathfrak{Z}_M$ can be written as
\begin{equation}
\mathfrak{Z}_M=\sum_{n_M}e^{{\beta}{\mu}n_M}\sum_{\{n_{M-1}\}}e^{\beta\epsilon_2\sum^*n_in_j}\prod_k^{\gamma}\mathfrak{z}_M\left(n_M,n_{M-1}^k\right),
\label{eq:ZM}
\end{equation}
where $\sum^*$ sums over all pairs $(i,j)$ of sites in generation $M-1$. The term $e^{{\beta}{\mu}n_M}$ takes into account the occupation of the center site, and the term $e^{\beta\epsilon_2\sum^*n_in_j}$ takes into account the interaction between different branches due to the 2NN interaction between the $\gamma$ sites in generation $M-1$. Fig.~\ref{fig:separate}(a) gives a graphical representation of Eq.~(\ref{eq:ZM}). In our solution, we explicitly account for 2NN interaction between branches of the same generation using the same approach as Ref.~\cite{katsura}. Such branch interactions do not appear in nearest neighbor only models, and many longer ranged models in the past have ignored branch interactions.

When $n$ of the $\gamma$ sites in generation $M-1$ are occupied, the interaction energy of those sites across branches of the same generation is $-\epsilon_2\binom{n}{2}$. There are $\binom{\gamma}{n}$ such configurations. Thus we can rewrite $\mathfrak{Z}_M$ as
\begin{equation}
\mathfrak{Z}_M=\sum_{n_M}e^{{\beta}{\mu}n_M}\sum_{n=0}^{\gamma}\binom{\gamma}{n}e^{\beta\epsilon_2\binom{n}{2}}\mathfrak{z}_M^{n}\left(n_M,1\right)\mathfrak{z}_M^{\gamma-n}\left(n_M,0\right).
\label{eq:ZM1}
\end{equation}

A branch $\mathcal{B}_m$ is composed of $\gamma-1$ sub-branches $\mathcal{B}_{m-1}$, so the partial grand partition function $\mathfrak{z}_m\left(n_m,n_{m-1}\right)$  can be written as 
\begin{widetext}
\begin{equation}
\mathfrak{z}_m\left(n_m,n_{m-1}\right)=e^{{\beta}{\mu}n_{m-1}}e^{{\beta}{\epsilon_1}n_mn_{m-1}}\sum_{\{n_{m-2}\}}e^{\beta\epsilon_2\left(n_m\sum^{'}n_i+\sum^{''}n_in_j\right)}\prod_k^{\gamma-1}\mathfrak{z}_{m-1}\left(n_{m-1},n_{m-2}^k\right),
\label{eq:zm0}
\end{equation}
\end{widetext}
where $\sum^{'}$ sums over all sites $i$ in generation $m-2$, and $\sum^{''}$ sums over all pairs $(i,j)$ of sites in generation $m-2$. The term $e^{{\beta}{\mu}n_{m-1}}$ takes into account the occupation of the site in generation $m-1$, the term $e^{{\beta}{\epsilon_1}n_mn_{m-1}}$ takes into account the NN interaction between the site in generation $m$ and the site in generations $m-1$, the term $e^{\beta\epsilon_2n_m\sum^{'}n_i}$ takes into account the 2NN interaction between the site in generation $m$ and sites in generations and $m-2$, and the term $e^{\beta\epsilon_2\sum^{''}n_in_j}$ takes into account the 2NN interaction across sites in generation $m-2$ of separate branches. Fig.~\ref{fig:separate}(b) gives an easier to follow graphical representation of Eq.~(\ref{eq:zm0}).

We can rewrite $\mathfrak{z}_m\left(n_m,n_{m-1}\right)$ using the same approach used for Eq.~(\ref{eq:ZM1}) as
\begin{widetext}
\begin{equation}
\mathfrak{z}_m\left(n_m,n_{m-1}\right)=e^{{\beta}{\mu}n_{m-1}}e^{{\beta}{\epsilon_1}n_mn_{m-1}}\sum_{n=0}^{\gamma-1}\binom{\gamma-1}{n}e^{\beta\epsilon_2\left(n_mn+\binom{n}{2}\right)}\mathfrak{z}_{m-1}^n\left(n_{m-1},1\right)\mathfrak{z}_{m-1}^{\gamma-1-n}\left(n_{m-1},0\right).
\label{eq:zm}
\end{equation}
\end{widetext}
This is the recursion relation for our model on the Bethe lattice, which facilitates analytic solution. We consider the possible values of $n_m$ and $n_{m-1}$ of Eq.~(\ref{eq:zm}), which can be simplified if we define the following
\begin{equation}
\label{eq:abcm}
a_m\equiv\frac{\mathfrak{z}_m\left(1,1\right)}{\mathfrak{z}_m\left(0,0\right)}\qquad b_m\equiv\frac{\mathfrak{z}_m\left(1,0\right)}{\mathfrak{z}_m\left(0,0\right)}\qquad c_m\equiv\frac{\mathfrak{z}_m\left(0,1\right)}{\mathfrak{z}_m\left(0,0\right)}.
\end{equation}

We evaluate Eq.~\eqref{eq:abcm} using the four cases for $n_m$ and $n_{m-1}$ in Eq.~\eqref{eq:zm}. In the limit $M\to\infty$, we obtain
\begin{equation}
a=\frac{e^{\beta\left(\mu+\epsilon_1\right)}\sum_{n=0}^{\gamma-1}\binom{\gamma-1}{n}e^{\beta\epsilon_2\binom{n+1}{2}}a^nb^{\gamma-1-n}}{\sum_{n=0}^{\gamma-1}\binom{\gamma-1}{n}e^{\beta\epsilon_2\binom{n}{2}}c^n}
\label{eq:a}
\end{equation}
\begin{equation}
b=\frac{\sum_{n=0}^{\gamma-1}\binom{\gamma-1}{n}e^{\beta\epsilon_2\binom{n+1}{2}}c^n}{\sum_{n=0}^{\gamma-1}\binom{\gamma-1}{n}e^{\beta\epsilon_2\binom{n}{2}}c^n}
\label{eq:b}
\end{equation}
\begin{equation}
c=\frac{e^{\beta\mu}\sum_{n=0}^{\gamma-1}\binom{\gamma-1}{n}e^{\beta\epsilon_2\binom{n}{2}}a^nb^{\gamma-1-n}}{\sum_{n=0}^{\gamma-1}\binom{\gamma-1}{n}e^{\beta\epsilon_2\binom{n}{2}}c^n},
\label{eq:c}
\end{equation}
where $a=\lim_{M \to \infty}a_M$, $b=\lim_{M\to \infty}b_M$, and $c=\lim_{M\to \infty}c_M$.

From Eq.~(\ref{eq:ZM1}) we can directly derive the expectation value of the occupancy of the center site
\begin{widetext}
\begin{equation}
<n_M>=\frac{e^{{\beta}{\mu}}\sum_{n=0}^{\gamma}\binom{\gamma}{n}e^{\beta\epsilon_2\binom{n}{2}}\mathfrak{z}_M^{n}\left(1,1\right)\mathfrak{z}_M^{\gamma-n}\left(1,0\right)}{\sum_{n_M}e^{{\beta}{\mu}n_M}\sum_{n=0}^{\gamma}\binom{\gamma}{n}e^{\beta\epsilon_2\binom{n}{2}}\mathfrak{z}_M^{n}\left(n_M,1\right)\mathfrak{z}_M^{\gamma-n}\left(n_M,0\right)},
\end{equation}
and in the limit of $M$ going to infinity
\begin{equation}
\label{eq:bethe-rho}
\rho=\lim_{M\to \infty}<n_M>=\frac{e^{{\beta}{\mu}}\sum_{n=0}^{\gamma}\binom{\gamma}{n}e^{\beta\epsilon_2\binom{n}{2}}a^nb^{\gamma-n}}{e^{{\beta}{\mu}}\sum_{n=0}^{\gamma}\binom{\gamma}{n}e^{\beta\epsilon_2\binom{n}{2}}a^nb^{\gamma-n}+\sum_{n=0}^{\gamma}\binom{\gamma}{n}e^{\beta\epsilon_2\binom{n}{2}}c^n}.
\end{equation}

The free energy is calculated using the method Gujrati proposed~\cite{gujrati}
\begin{equation}
\label{eq:gujrati}
\frac{P}{kT}=\frac{1}{2}\lim_{M \to \infty}\left[\ln \mathfrak{Z}_M-\prod_{k=1}^{\gamma-1}\ln \mathfrak{Z}_{M-1}\right].
\end{equation}
\end{widetext}
Note that this expression was derived by Gujrati in the specific case of nearest-neighbor interactions only. With the graphical representations in Fig.~\ref{fig:separate}, we can show that this expression is also valid for models with 2NN interactions (if interactions across separate branches are not ignored). Specifically, we can divide the whole tree $\mathcal{T}_M$ into $\gamma$ branches $\mathcal{B}_M$, and further divide them into $\gamma\times(\gamma-1)$ shorter branches $\mathcal{B}_{M-1}$. In the graphical representation we get $\gamma+1$ triangles and $\gamma$ lines in this process. We can divide the $\gamma-1$ shorter trees $\mathcal{T}_{M-1}$ into $\gamma\times(\gamma-1)$ shorter branches $\mathcal{B}_{M-1}$. In the graphical representation we get $\gamma-1$ triangles and $\gamma-1$ lines in this process . The difference is $2$ triangles and 1 line, which represent the occupations of 2 lattice sites, $\gamma\times(\gamma-1)$ 2NN interactions, and one NN interaction. This is the free energy of two sites. Thus we can divide it by two to obtain the free energy per site ({\it i.e} pressure P).

Eq.~(\ref{eq:gujrati}) can be simplified by substituting $\mathfrak{Z}_M$ and $\mathfrak{Z}_{M-1}$ using Eq.~(\ref{eq:ZM1}), and factoring out powers of $\mathfrak{z}_M(0,0)$ and $\mathfrak{z}_{M-1}(0,0)$ to express the rest with the ratios $a_M$, $b_M$, $c_M$, $a_{M-1}$, $b_{M-1}$, $c_{M-1}$. Further substituting $\mathfrak{z}_M(0,0)$ using Eq.~(\ref{eq:zm}), we can cancel out the powers of $\mathfrak{z}_{M-1}(0,0)$ and express the right hand side of Eq.~(\ref{eq:gujrati}) entirely with the ratios $a_M$, $b_M$, $c_M$ and $a_{M-1}$, $b_{M-1}$, $c_{M-1}$. When we take the limit $M\to\infty$, we obtain

\begin{widetext}
\begin{equation}
\label{eq:bethe-P}
\frac{P}{kT}=\frac{\gamma}{2}\ln \sum_{n=0}^{\gamma-1}\binom{\gamma-1}{n}e^{\beta\epsilon_2\binom{n}{2}}c^n-\frac{\gamma-2}{2}\ln\left[e^{\beta\mu}\sum_{n=0}^{\gamma}\binom{\gamma}{n}e^{\beta\epsilon_2\binom{n}{2}}a^nb^{\gamma-n}+\sum_{n=0}^{\gamma}\binom{\gamma}{n}e^{\beta\epsilon_2\binom{n}{2}}c^n\right].
\end{equation}
\end{widetext}

We have now derived everything we need to evaluate the phase behavior. To evaluate $\rho$ and $P$ for a given $\mu$ and $T$, we first obtain $a$, $b$, $c$ by solving the system of implicit equations \eqref{eq:a,eq:b,eq:c} using the iterative scheme described in Sec.~\ref{method-MF}, and use these values to calculate density $\rho$ with Eq.~(\ref{eq:bethe-rho}) and pressure $P$ with Eq.~(\ref{eq:bethe-P}). Similarly we repeat the iteration procedure with different initial $a$, $b$, and $c$ to search for possible stable and metastable states. When multiple solutions are found, we take the solution with the highest pressure (lowest free energy).

\subsection{Monte Carlo Simulations}
\label{method-MC}

To evaluate the exact phase diagrams, we perform Monte Carlo (MC)
simulations in the grand canonical ensemble (fixed $\mu$, $V$, and $T$)
-- or GCMC for short~\cite{frenkelBook}. For the 2D square lattice, our
system size is $40\times40$ (1600 sites). For the cubic and diamond
lattices, our system size is $12\times12\times12$ (1728 sites); for the
diamond lattice this corresponds to $6\times6\times6$ unit
cells. Periodic boundary conditions are implemented in all cases.

We first locate the rough location of critical points by the appearance
of a bimodal density distribution. To obtain accurate $T_c$, $\mu_c$,
and $\rho_c$, we use the fact that the model is expected to be in the
Ising universality class~\cite{db03}.  Specifically, we study the order
parameter $M=\rho-su$ and use the histogram re-weighting
technique~\cite{wilding} to make minor adjustments in $T$ and $\mu$ so
that the order parameter distribution $P(M)$ approaches that of the Ising
universality class. A detailed description of this procedure is given by
Refs.~\cite{wilding}.

To evaluate the phase boundaries in the subcritical region, we perform a series of GCMC simulations with multi-canonical biased sampling~\cite{frenkelBook,wilding} to allow us to efficiently sample both phases in a single simulation. To estimate the coexistence densities, we again apply histogram re-weighting to make minor adjustments in $T$ and $\mu$ so that the distribution $P(M)$ has same height for the two peaks corresponding to the two phases, and that the integral under each of the peaks, that is the probability for each phase, is equal.

Data used to build the histograms are obtained from pairs of $(N,E)$ values taken from 8 independent simulations, each running $5\times10^6$ MC steps per site. The $(N,E)$ data are taken every 10 MC steps for the entire lattice after an equilibrium of $5\times10^5$ MC steps per site.


\section{Results and Discussion}
\label{result}

\subsection{Symmetric Lattice Model}
\label{result-sym}

\begin{figure}
\centerline{
\includegraphics[width=1.75in]{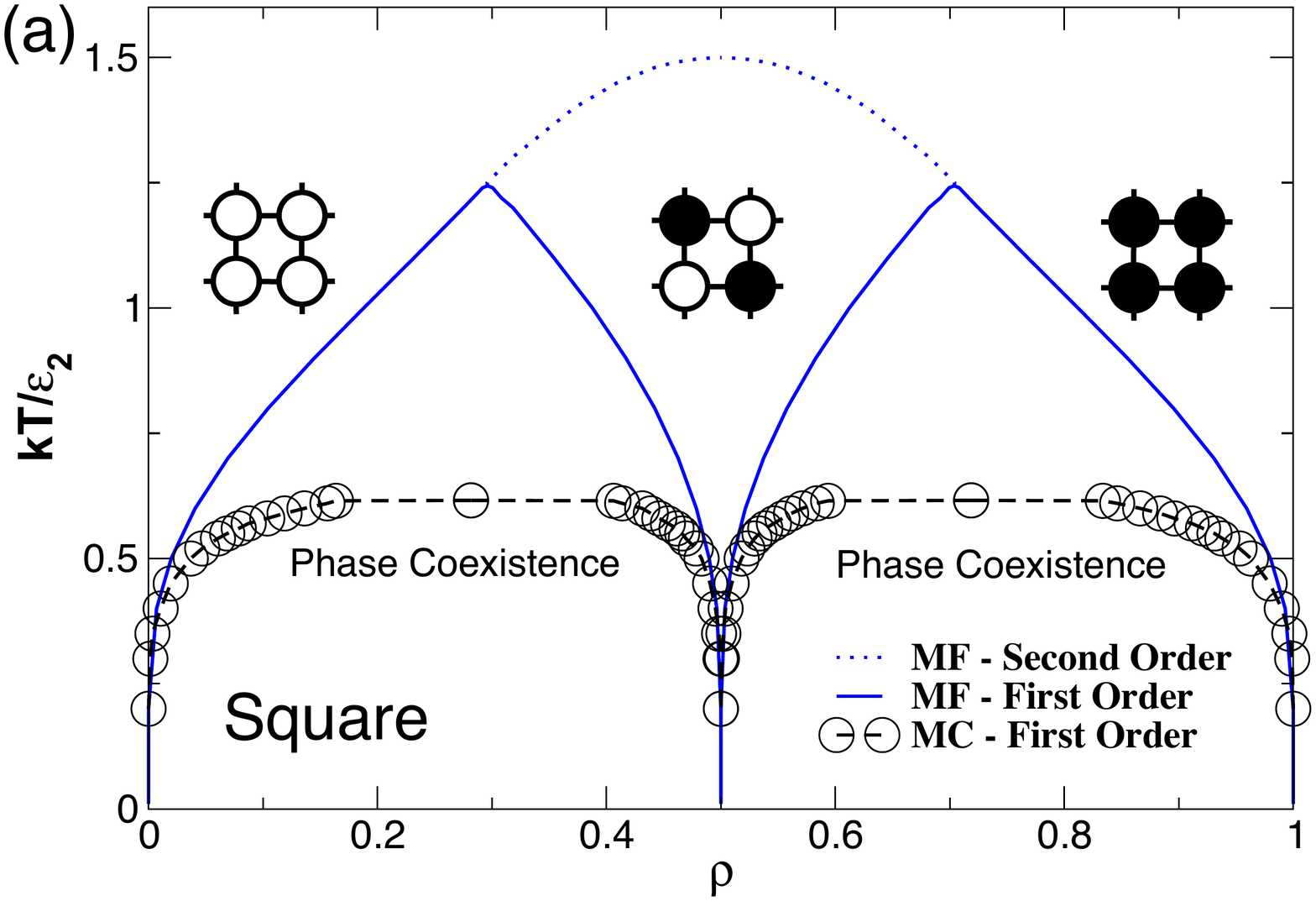}
\includegraphics[width=1.75in]{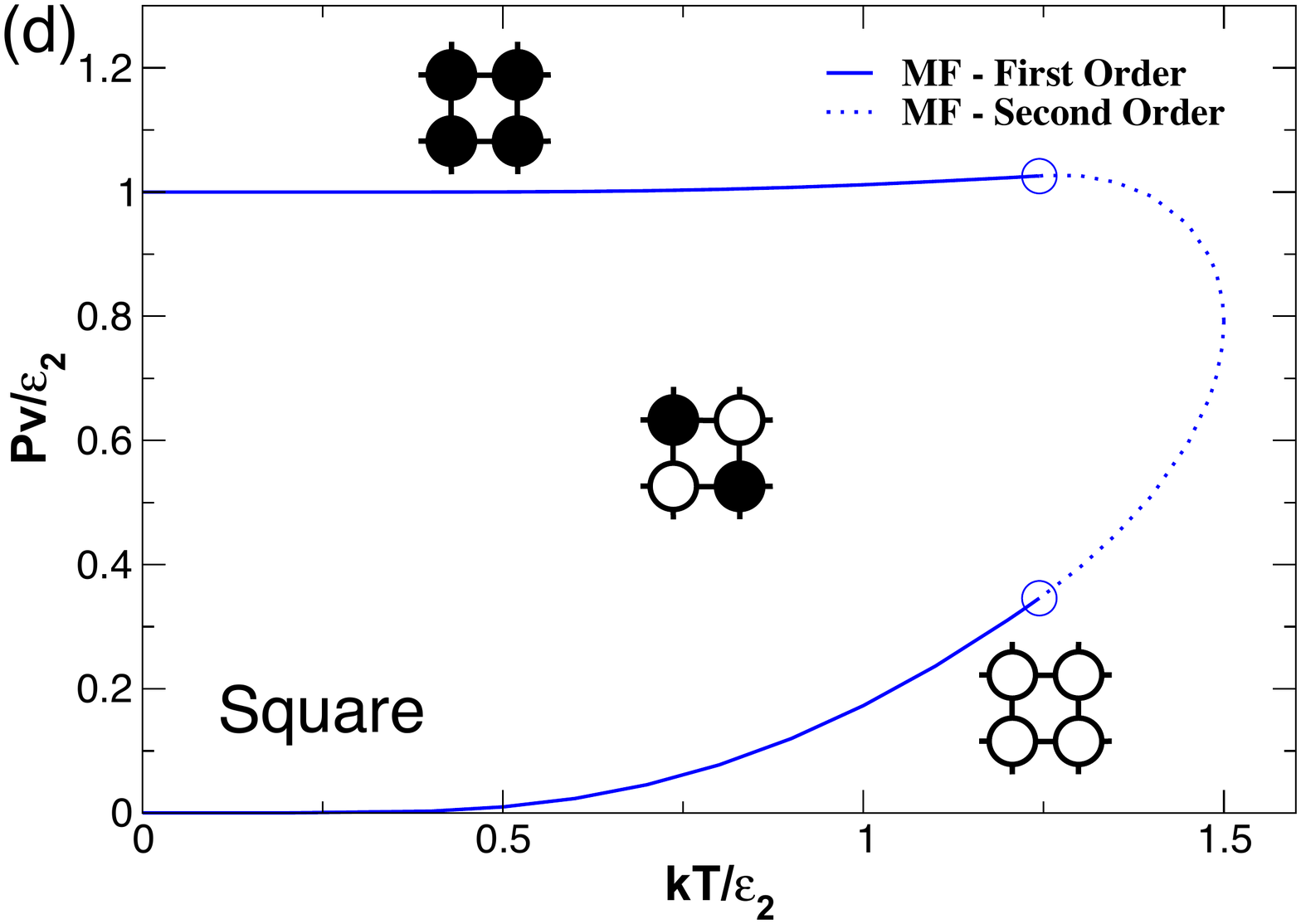}
}
\centerline{
\includegraphics[width=1.75in]{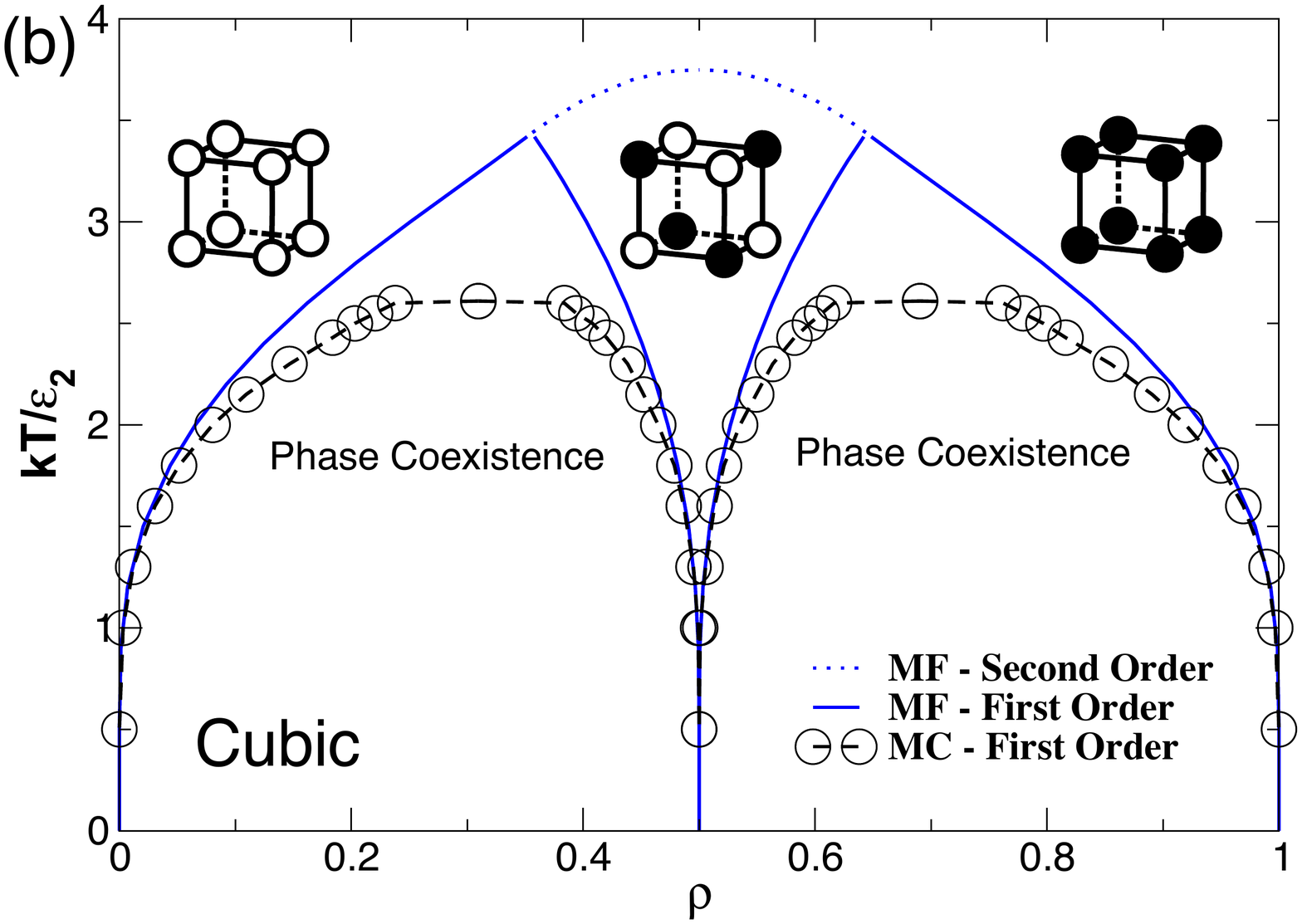}
\includegraphics[width=1.75in]{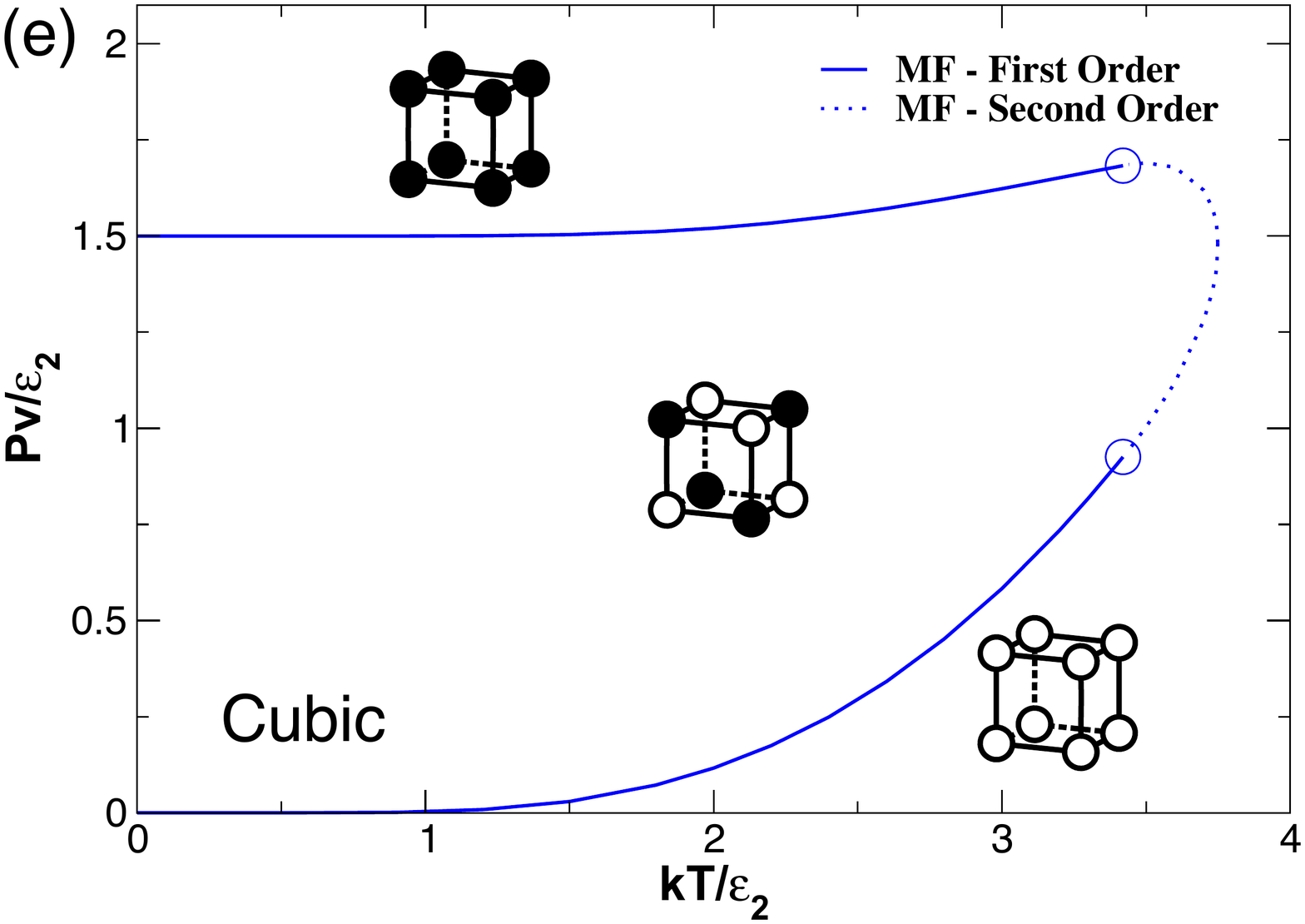}
}
\centerline{
\includegraphics[width=1.75in]{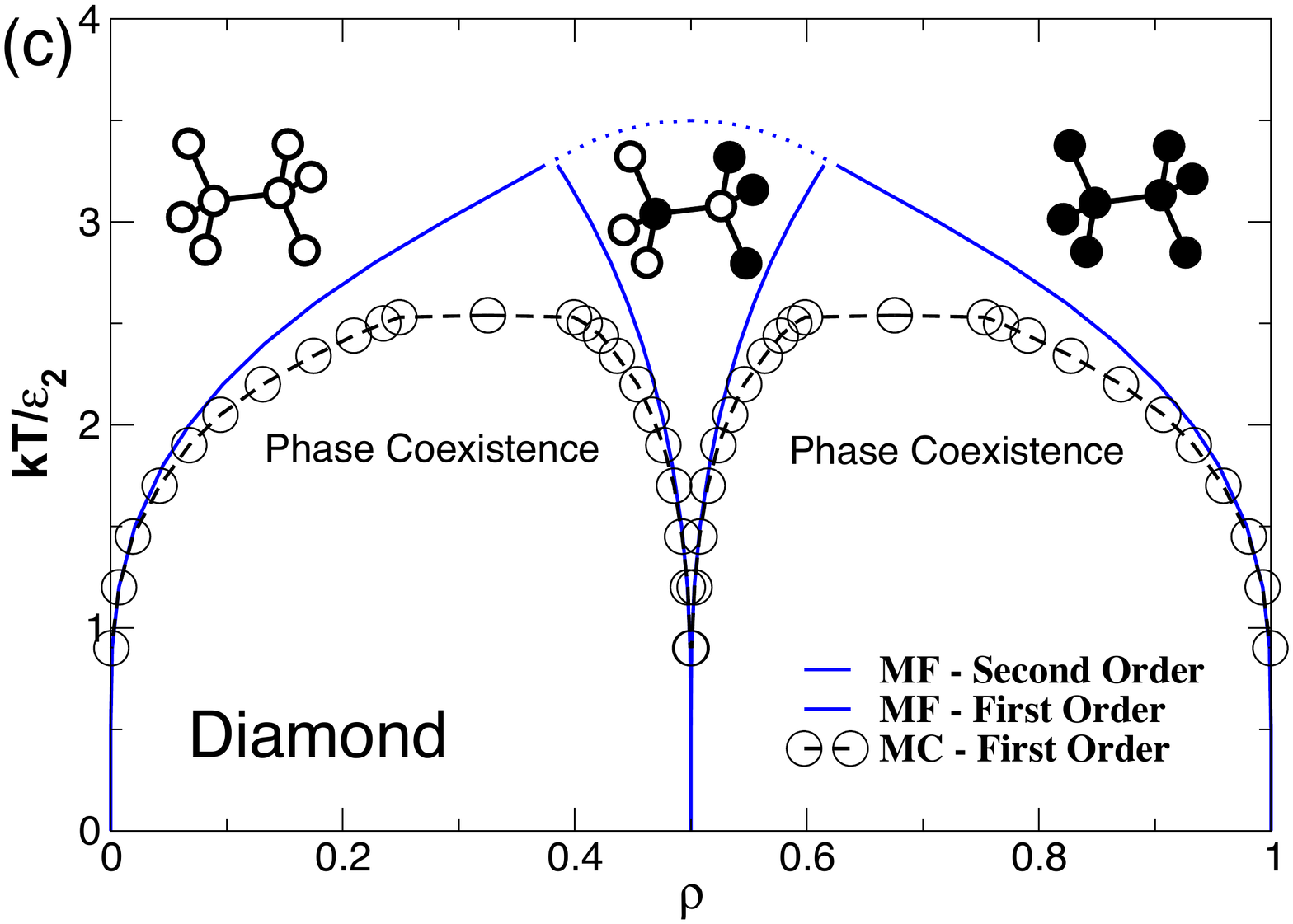}
\includegraphics[width=1.75in]{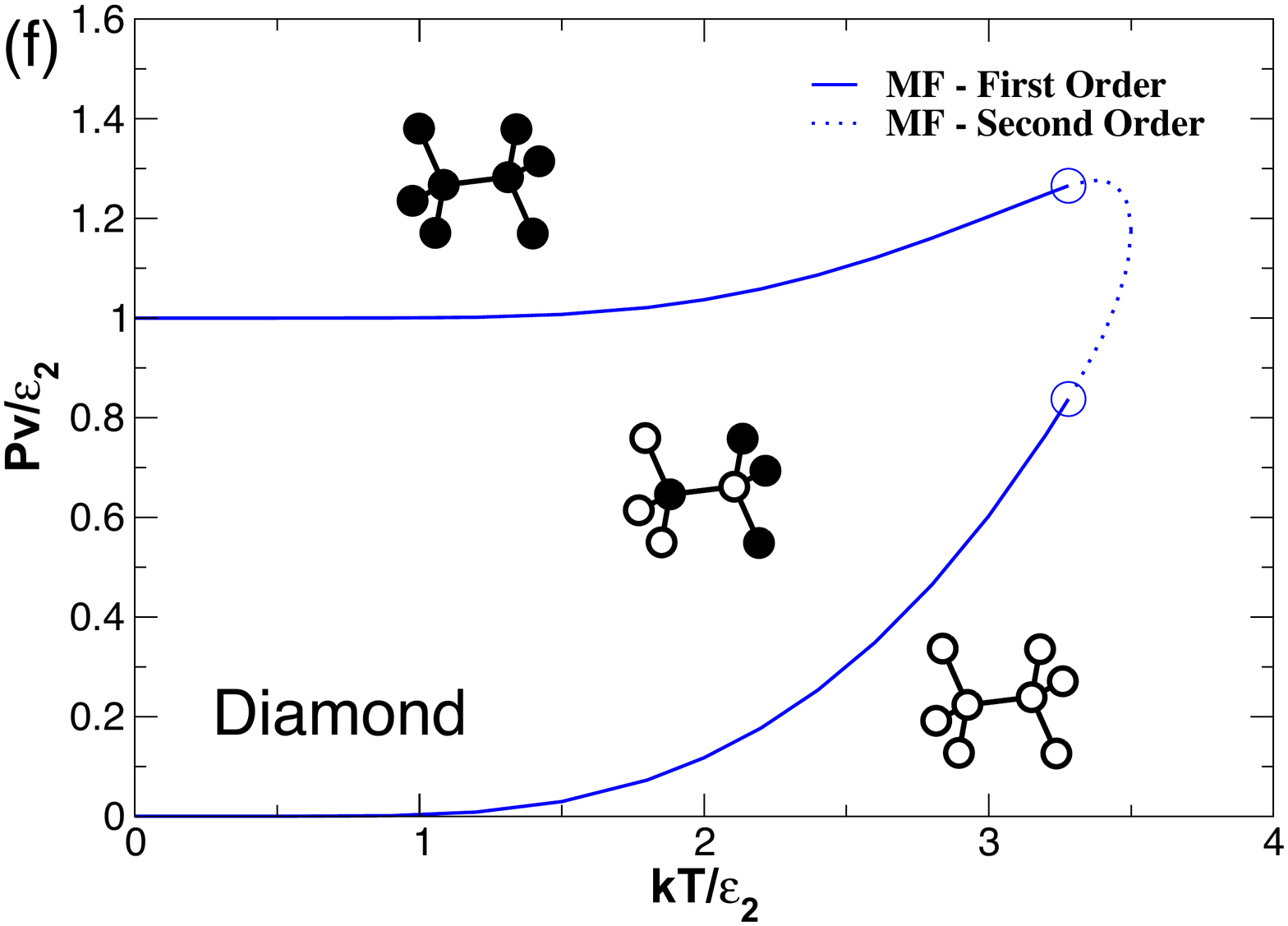}
}
\caption{Phase diagrams from the mean-field approximation and Monte Carlo simulations of the symmetric lattice model with $R=-1/2$. (a,d) square lattice. (b,e) cubic lattice. (c,f) diamond lattice.}
\label{fig:phase-sym}
\end{figure}

To illustrate how the long-ranged attractions of the model Hamiltonian Eq.~\eqref{eq:sym2NN} affect phase behavior, we focus on the phase diagram for the case of $R=-1/2$, {\it i.e.} second neighbor attractions twice that of first neighbor repulsion. We show the resulting $\rho$-$T$ phase diagrams for the mean-field approximation and for MC simulations on square, cubic, and diamond lattices in Fig.~\ref{fig:phase-sym}(a,b,c). On all three lattices we find three thermodynamically distinct phases analogous to: (i) unassociated molecules of gas, (ii) liquid I, a single network of alternatingly filled sites, and (iii) liquid II, a double interpenetrating network with all sites occupied. Hence, our system demonstrates that long-ranged attraction and short-ranged repulsion alone can generate multiple high-density phases via interpenetration. This is precisely the mechanism proposed for DNA functionalized nanoparticles~\cite{hlss}.

By construction of the model, the $\rho$-$T$ phase diagrams are
symmetric about $\rho$=0.5. The mean-field approximation correctly
predicts the number of transitions and the qualitative shape of the
first-order transition boundaries. As expected, the mean-field
approximation overestimates the terminal temperature of the first order
transition and incorrectly predicts a second-order phase
transition. While the pressure is not readily available from the MC
simulations, Eq.~\eqref{eq:MF-sym-P} allows us to readily calculate the
$T$-$P$ phase diagram for each lattice
(Fig.~\ref{fig:phase-sym}(d,e,f)). In all cases, the slope of the
coexistence lines in the $T$-$P$ plane are positive, unlike the slope of
the coexistence lines in water. For the coexistence lines, the slope
$\left(\frac{\partial P}{\partial T}\right)_\mu=\frac{\Delta s}{\Delta
  v}$ (the Clausius-Clapeyron relation)~\cite{HESbook}; thus,
interpenetration alone does not require an anomalous relation between
the difference in entropy $\Delta s$ and difference in volume $\Delta v$
of the two phases.


\begin{figure}
\centerline{
\includegraphics[width=1.75in]{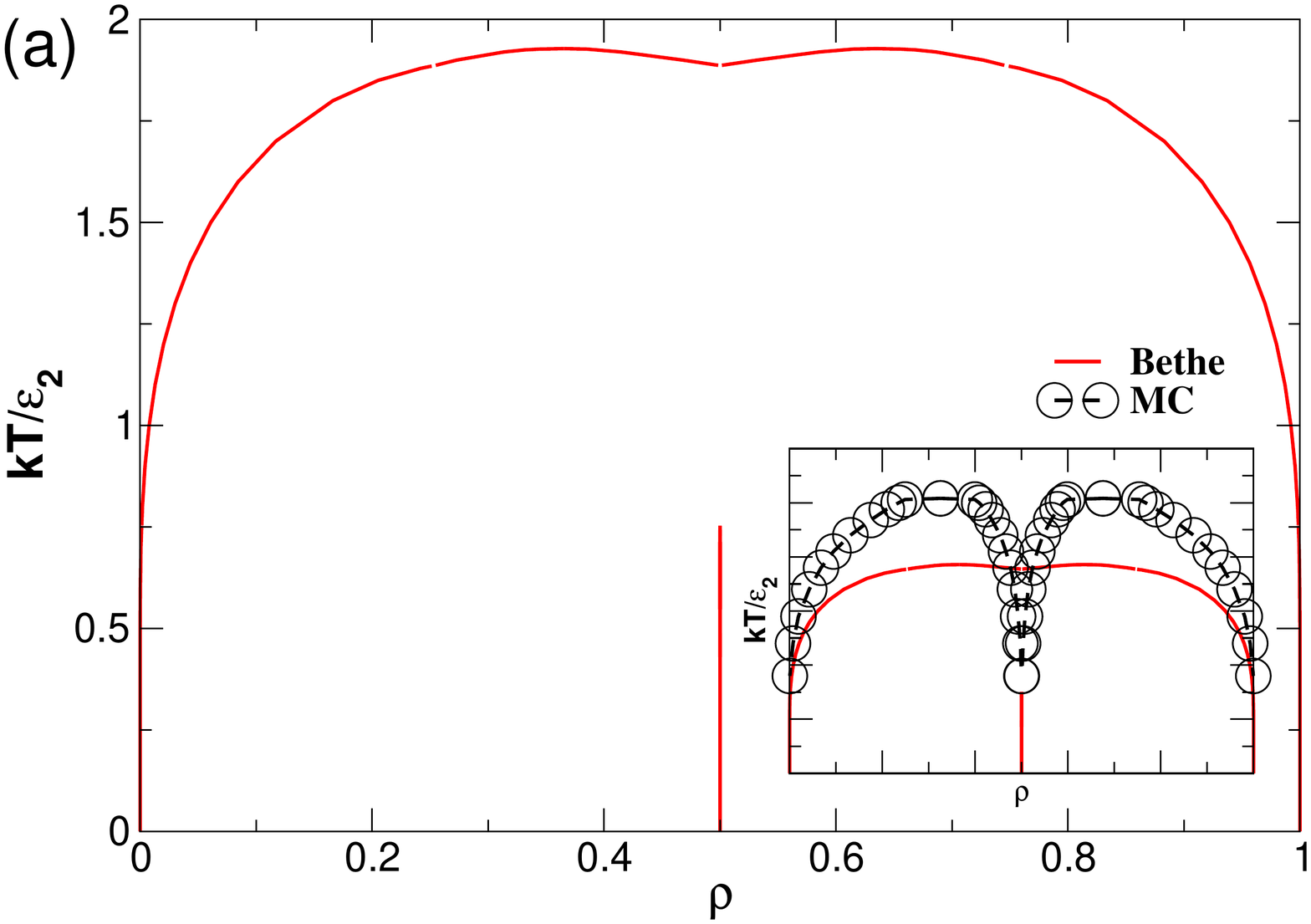}
\includegraphics[width=1.75in]{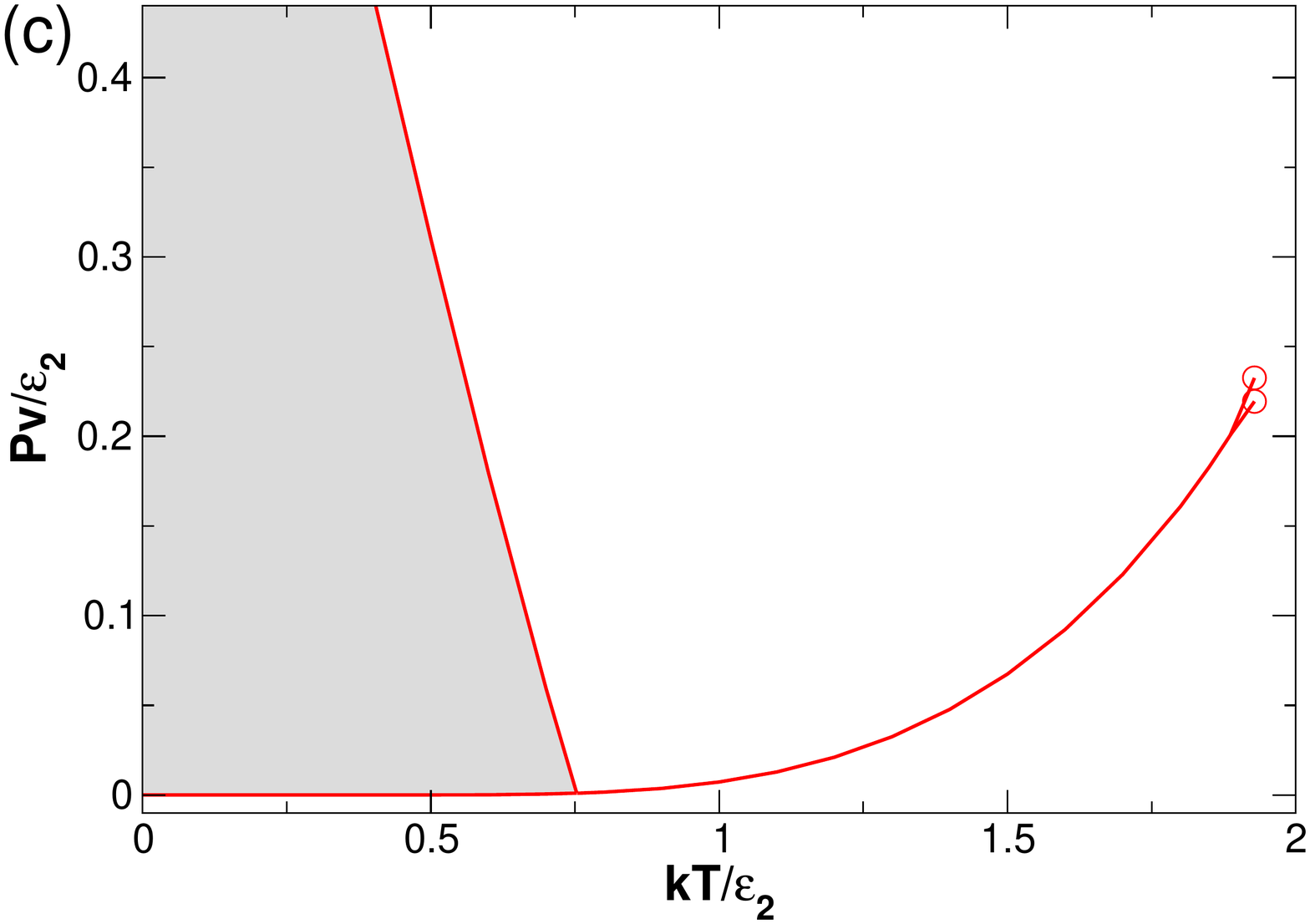}
}
\centerline{
\includegraphics[width=1.75in]{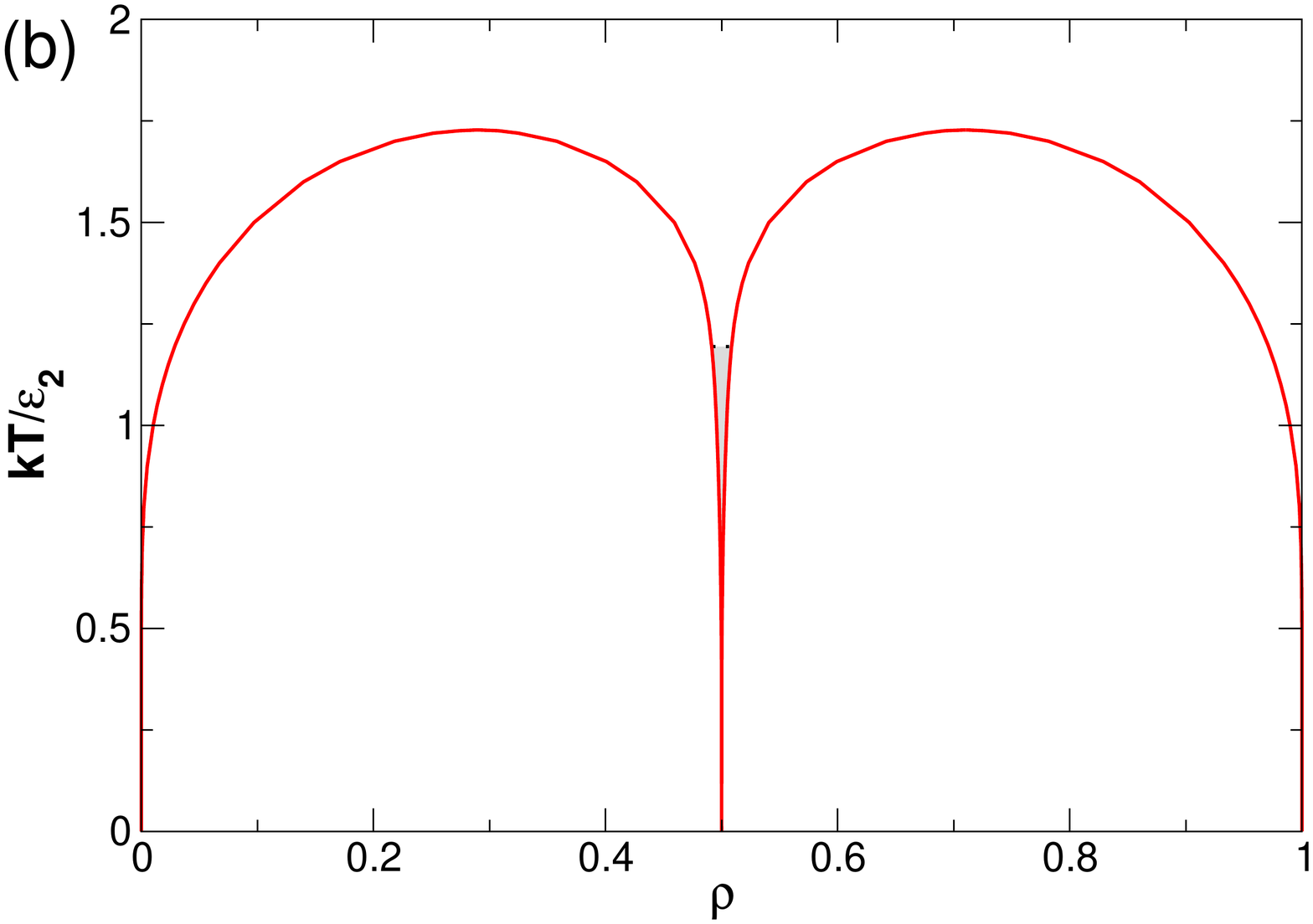}
\includegraphics[width=1.75in]{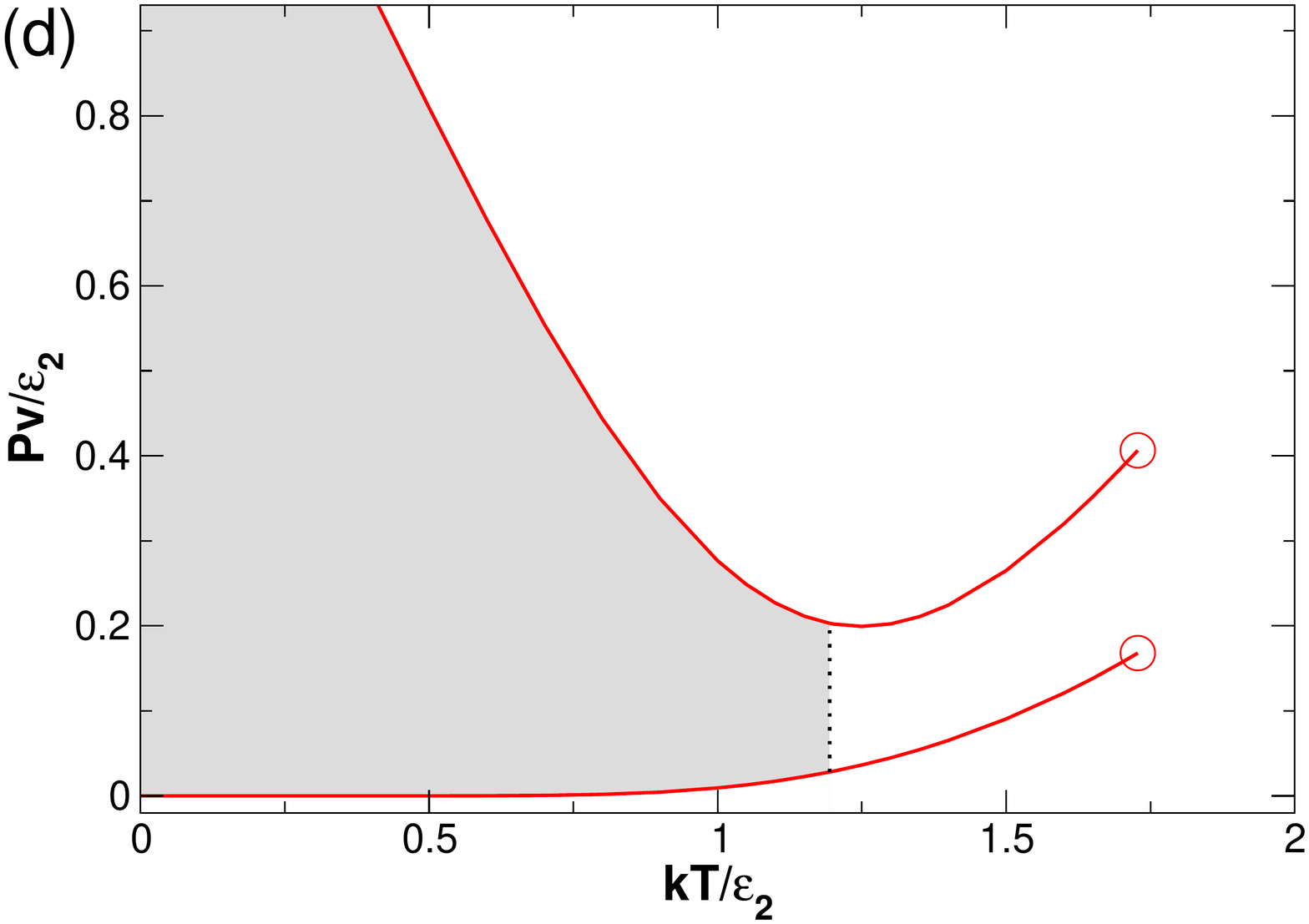}
}
\caption{Phase diagrams of the symmetric model on Bethe lattice, calculated from exact solution. (a,c) $R=-1/2$. (b,d) $R=-3/4$. The shaded regions indicate the location of anomalous $\rho$ dependence, {\it i.e.} $\alpha_P<0$.}
\label{fig:phase-bethe}
\end{figure}

Since the Bethe lattice geometry is distinct ({\it i.e.} no closed
loops) from those studied for the MF and MC solutions, we present the
results separately. To mimic the behavior of the diamond lattice studied
in Fig.~\ref{fig:phase-sym}(c,f) and known to occur in many polyamorphic
fluids, we examine the case $\gamma=4$
(Fig.~\ref{fig:phase-bethe}). Moreover, both the Bethe lattice with
$\gamma=4$ and the diamond lattice have 12 second neighbors, so the
results can be expected to be similar. To our surprise, the phase
behavior for $R=-1/2$ is qualitatively different from that of the
diamond lattice. Specifically, the coexistence lines merge for
$1.754<kT/\epsilon_2<1.886$, resulting in two triple points in the
$T$-$P$ phase diagram. In other words, there are only liquid and gas
states for $1.754<kT/\epsilon_2<1.886$, but just below and above there
are three distinct states. We also study $R=-3/4$; in this case, there
are two distinct phase transitions that do not merge.

\begin{figure}
\begin{center}
\includegraphics[width=3.5in]{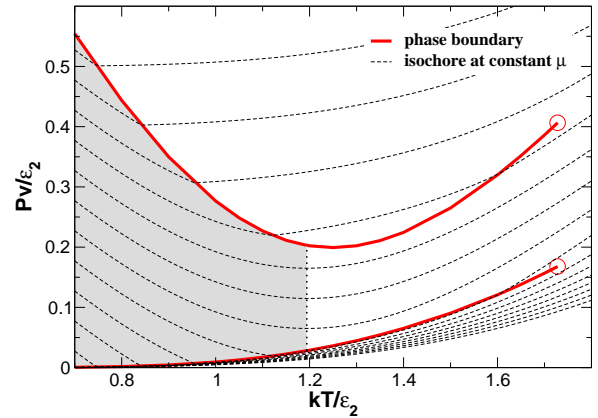}
\end{center}
\caption{Isochores of $P$ at constant $\mu$ for the Bethe lattice $R=-3/4$ case. Density anomaly exists for regions where the slope of the isochores are negative; such regions are shaded. The slope of the isochores between the two boundaries all go to zero at $kT/\epsilon_2=1.194$, as indicated by the dotted line. }
\label{fig:bethe-isochem}
\end{figure}

The presence of a negatively sloped coexistence line in the $T$-$P$
plane indicates an anomalous ratio $\frac{\Delta s}{\Delta v}<0$. This
suggest there may be anomalous density dependence, {\it i.e.} that the
isobaric expansivity $\alpha_P=-\frac{1}{\rho}\left(\frac{\partial
    \rho}{\partial T}\right)_{P,\mu}<0$. Since our solution does not
easily allow us to hold pressure fixed, we instead check the thermal
pressure coefficient $\gamma_V=\left(\frac{\partial P}{\partial
    T}\right)_{V,\mu}$, since $\gamma_V=\frac{\alpha_P}{\kappa_T}$,
where $\kappa_T$ is the isothermal compressibility. The sign of
$\alpha_P$ is determined by the sign of $\gamma_V$ since
$\kappa_T\geq0$. We find that for the $R=-1/2$ case, the density is
anomalous throughout the intermediate
phase. Fig.~\ref{fig:bethe-isochem} shows that for the $R=-3/4$ case,
the density is anomalous for $kT/\epsilon_2<1.194$ in the intermediate
phase. In the phase diagrams in Fig.~\ref{fig:phase-bethe}, density
anomaly regions are indicated by the shading.

\subsection{Asymmetric and Third-neighbor Lattice Models}
\label{result-asym-3NN}

\begin{figure}
\centerline{
\includegraphics[width=1.75in]{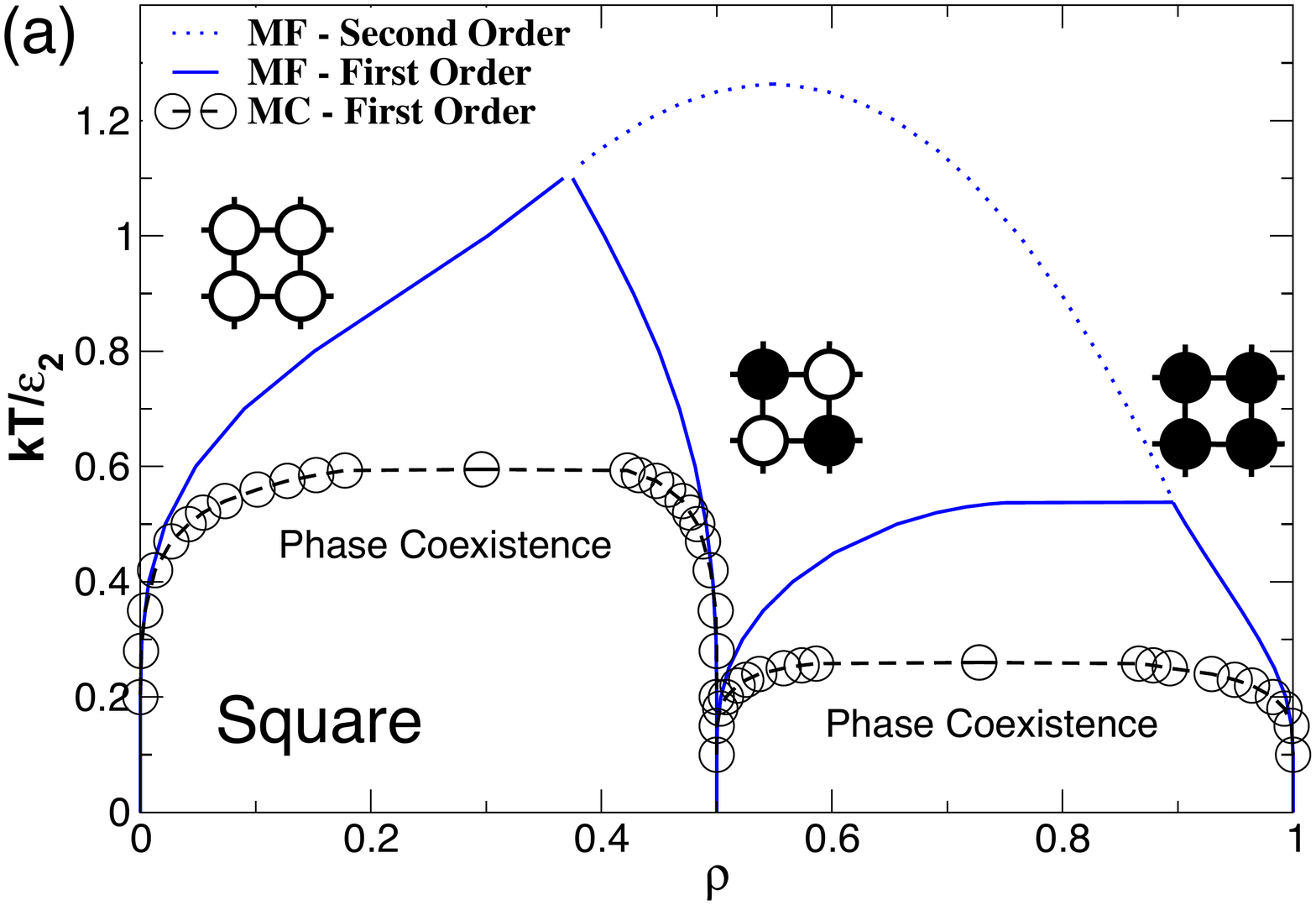}
\includegraphics[width=1.75in]{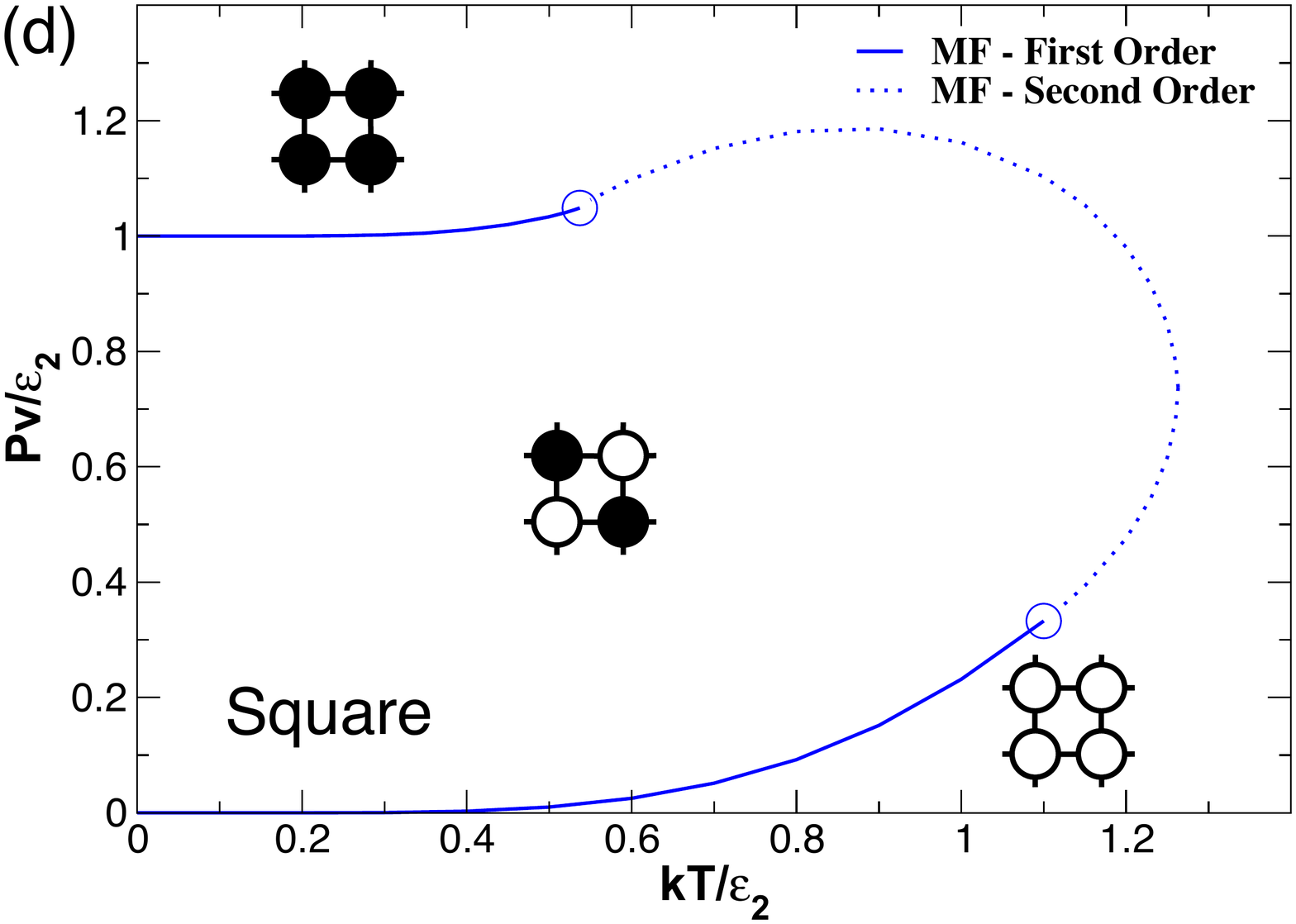}
}
\centerline{
\includegraphics[width=1.75in]{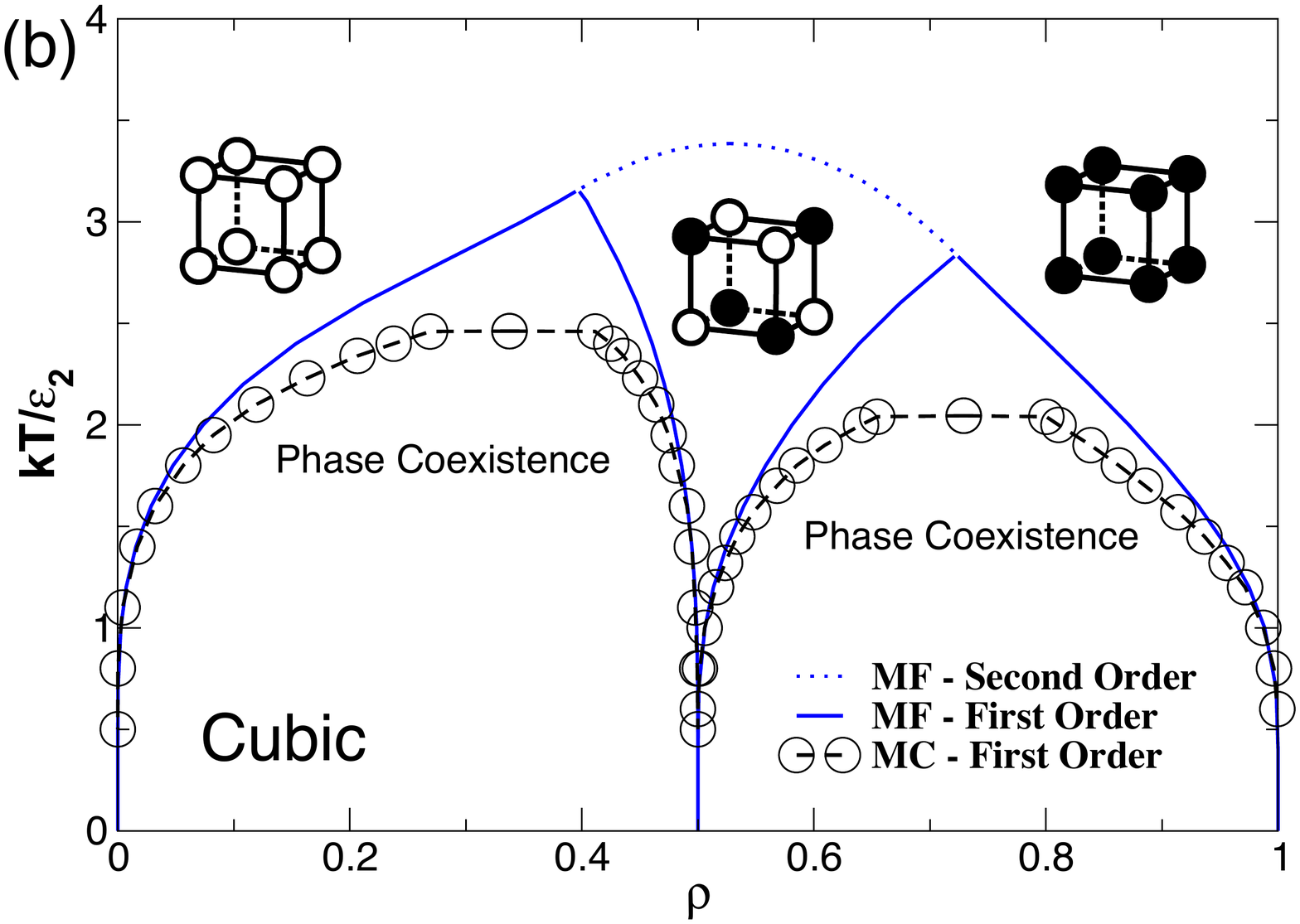}
\includegraphics[width=1.75in]{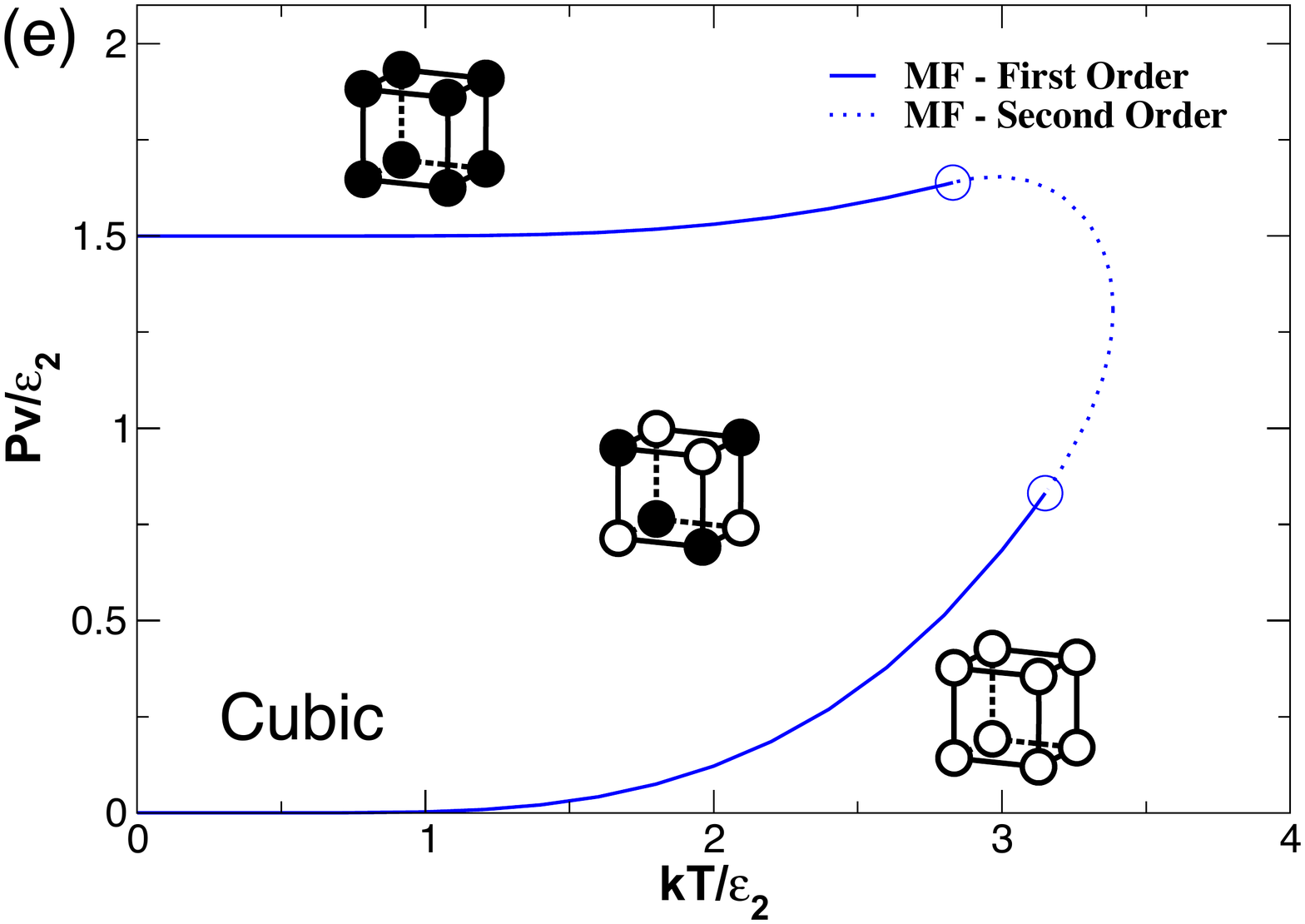}
}
\centerline{
\includegraphics[width=1.75in]{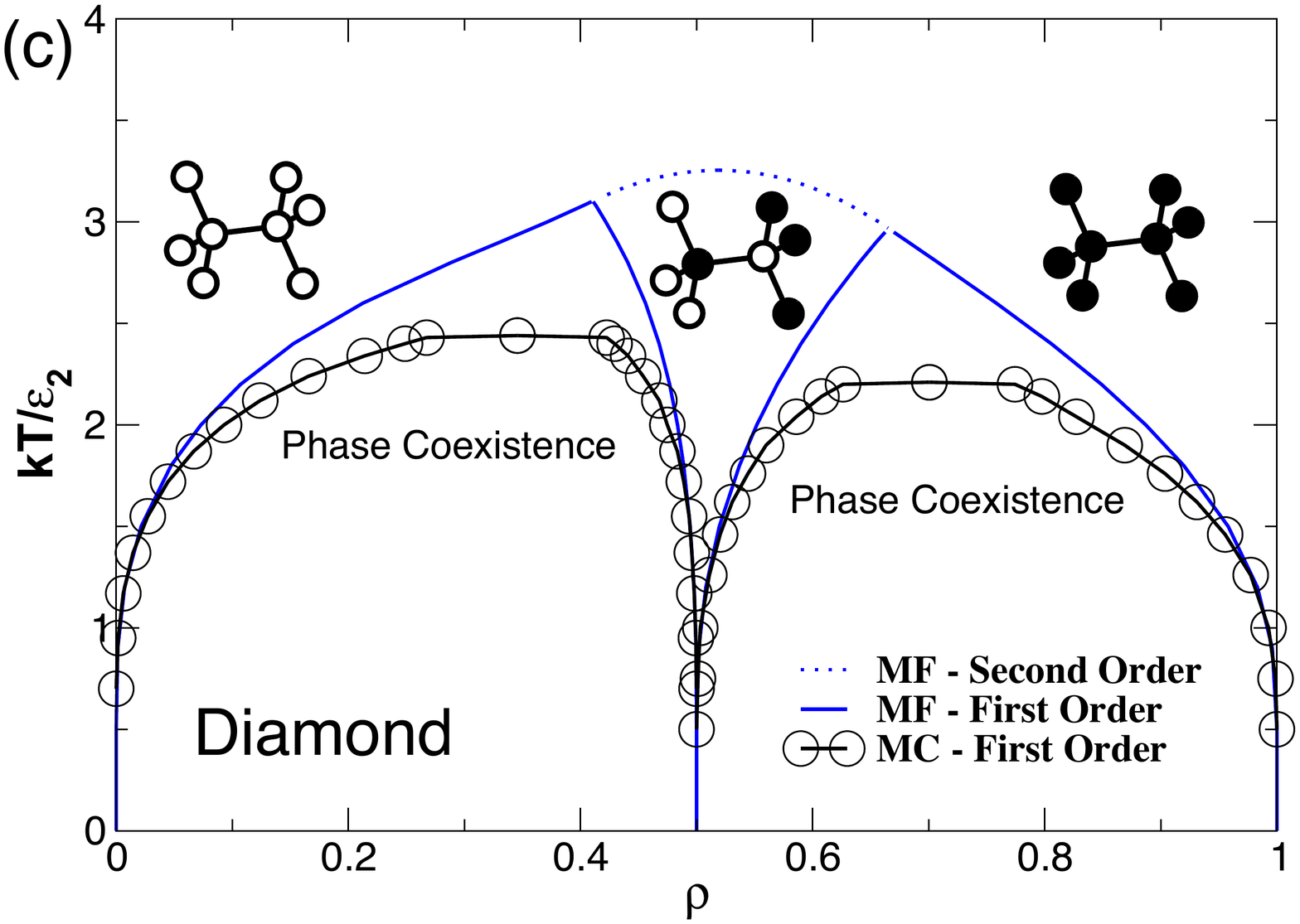}
\includegraphics[width=1.75in]{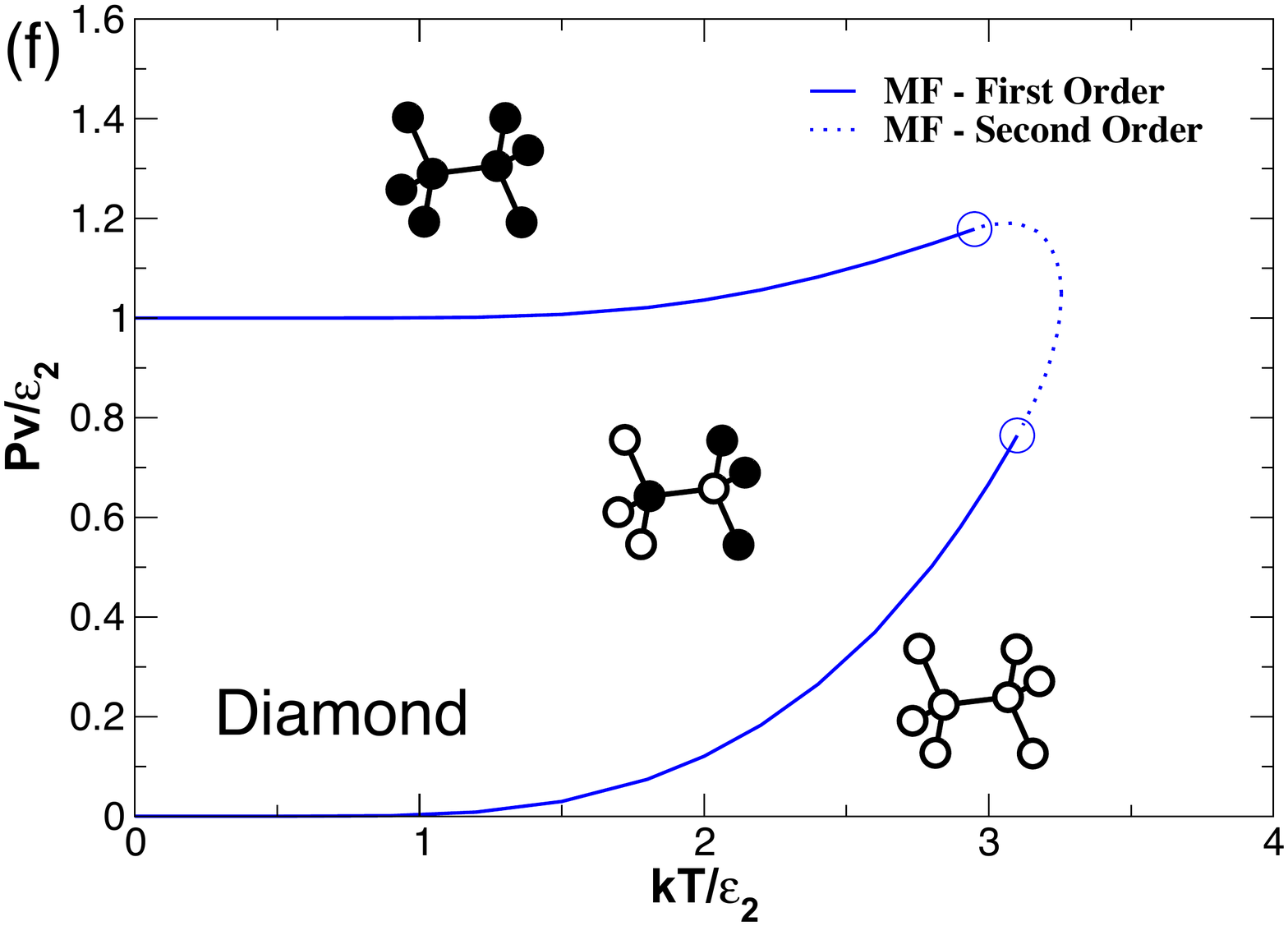}
}
\caption{Phase diagrams from the mean-field approximation and Monte Carlo simulations of the asymmetric lattice model with $R=-1/2$. (a,d) square lattice. (b,e) cubic lattice. (c,f) diamond lattice.}
\label{fig:phase-asym}
\end{figure}

We next examine the changes in the phase diagrams of the symmetric model (except for the Bethe lattice) when we include a 3-body interaction with first neighbor sites. For the same value $R=-1/2$, Fig.~\ref{fig:phase-asym} shows that the three-body term breaks the symmetry in $\rho$-$T$ phase diagrams. While the width of the transitions is unchanged, the high density critical point has a lower critical temperature than the low density critical point. The depression of the second critical temperature is consistent with the observation in polyamorphous systems that the high density critical point occurs at lower $T$ than the liquid-gas critical point. Aside from the difference in $T_c$, the phase diagrams in both $T$-$P$ and $\rho$-$T$ are qualitatively comparable.

\begin{figure}
\centerline{
\includegraphics[width=1.75in]{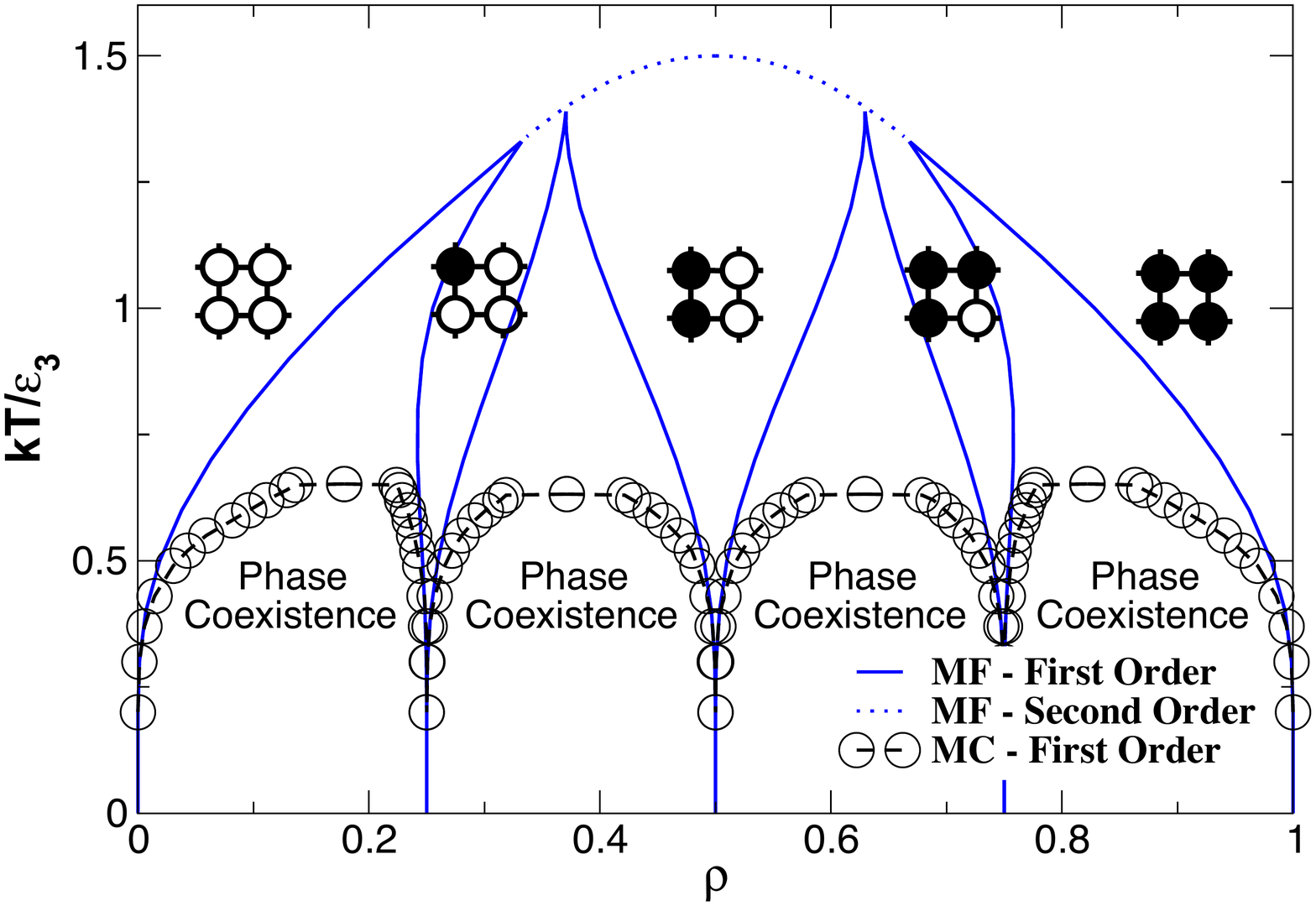}
\includegraphics[width=1.75in]{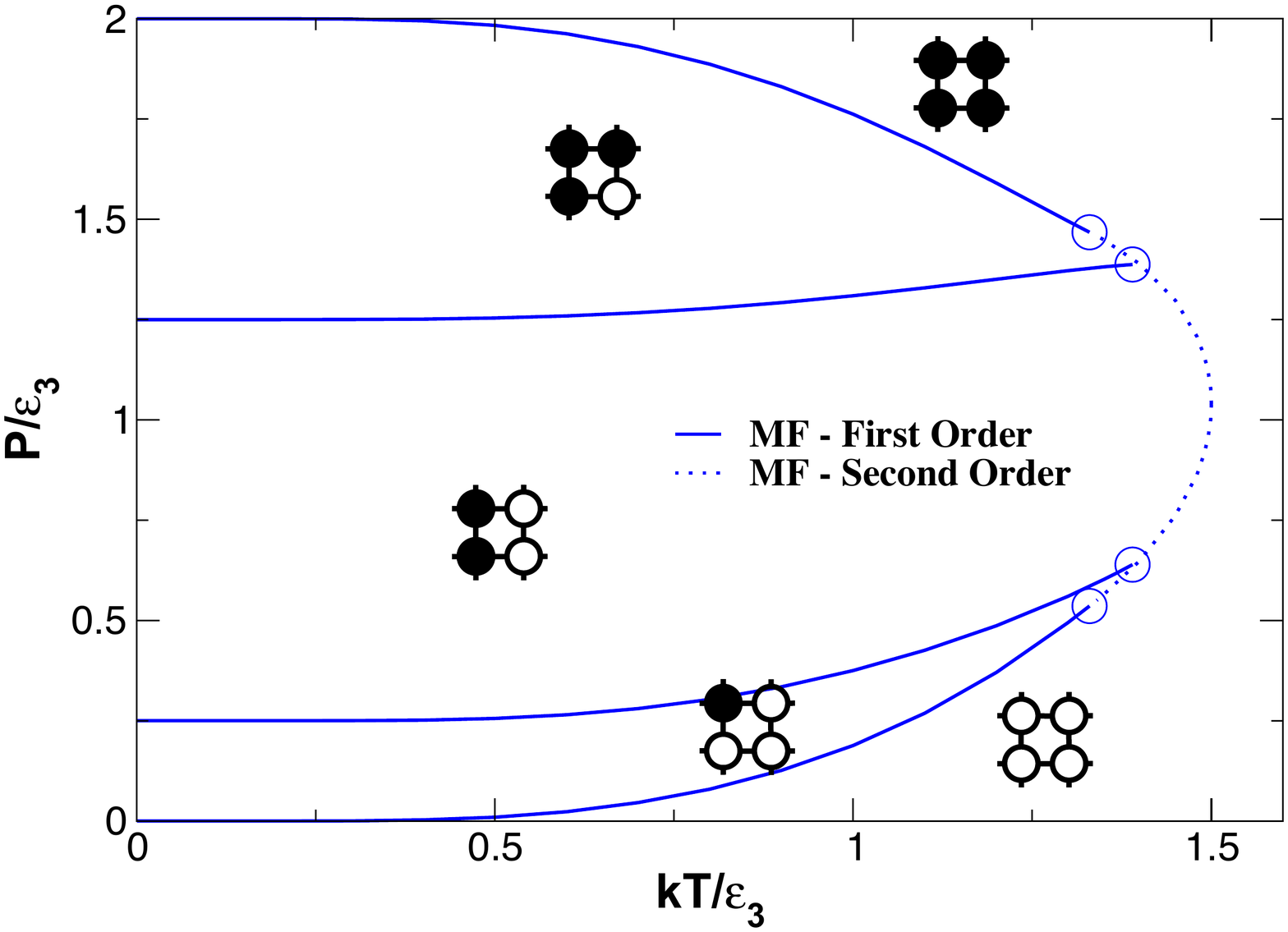}
}
\caption{Phase diagrams of the 3NN interaction model from the mean-field approximation and Monte Carlo simulations, with $R_1=R_2=-1/2$. Done on a 2D square lattice.} 
\label{fig:phase-3NN}
\end{figure}

The third-neighbor lattice model has a longer range for bonding, which opens more nearby sites that can be occupied by molecules in separate interpenetrating sub-lattices. Our MF and MC results for the phase behavior of the case $R_1=R_2=-1/2$ (Fig.~\ref{fig:phase-3NN}) confirm that the additional open sites allow for an even greater number of coexisting phases; specifically, we find up to five coexisting phases -- more than previously observed in any molecular systems. These five phases correspond to (i) gas, empty lattice, (ii) a single network of one sublattice, (iii) two networks with two sublattices occupied, (iv) three networks with three sublattices occupied, and (v) four networks with the whole lattice occupied. Although the coexistence line of the highest density transition in the T-P plane indicates an anomalous ratio $\frac{\Delta s}{\Delta v}<0$, by examining $\gamma_V$, we find there is no density anomaly.

The middle phase consists of sites occupied alternately in parallel (superantiferromagnetic) instead of sites occupied alternately in diagonal (antiferromagnetic). This can be explained by the ground state free energy of these two configurations. For an occupied site on a fully superantiferromagnetic configuration, half of the NN sites are occupied and none of the 2NN sites are occupied. For an occupied site on a fully antiferromagnetic configuration, none of the NN sites are occupied and all of the 2NN sites are occupied. For $R_1=R_2=-1/2$, the superantiferromagnetic configuration has less NN and 2NN repulsion, thus it has lower free energy. We can expect that if the NN repulsion is sufficiently stronger ($R_1<2R_2$), the stable phase in the middle would be antiferromagnetic.



\section{Conclusion}
\label{conclusion}

In the same spirit that the nearest neighbor lattice gas helps to understand the liquid-gas transition, we have used lattice models with an extended bonding range to understand how the formation of interpenetrating open networks can give rise to liquid-liquid transitions. The specific mechanism of interpenetration seems most applicable to recently studied DNA functionalized nanoparticles, where the range of bonding can be quite large compared with the core exclusion~\cite{hlss}.  

The presence of lattice sites in the lattice models we have studied provides a predetermined regular framework for the interpenetration. As a result, it is natural to generate distinct phases on the sublattices defined by second neighbors (or more complex sublattices in the case of the third neighbor model). However, in continuum systems, there is no such underlying lattice structure. For systems with highly directional interactions, a network structure may appear naturally. Interpenetration for tetrahedral networks is facilitated by the fact that the empty spaces of the tetrahedral network make a complementary tetrahedral network; the same is true for cubic network. For such systems, the free energy of distinct networks can be significantly lower than that of a distorted, connected structure. As a result, the system will preferentially phase segregate at densities where distinct networks with few defects are not possible. Thus, for networked structures lacking such symmetry, the interpenetration could be frustrated, thereby eliminating the free energy gap between distorted networks and distinct interpenetrating networks. Without such a free energy gap, no phase separation will occur. In particular, spherically symmetric systems with a long-ranged bonding term would likely not result in multiple phases due to interpenetration. However, packing constraints may still give rise to polyamorphic behavior in step potentials with carefully chosen step sizes~\cite{fmsbs,bs,sbfms}.

The lattice models we have considered might be made more specific to the problem of DNA functionalized nanoparticles by including specific interactions that could mimic the molecular recognition of DNA, and by controlling the number of bonds that a given site can participate in.  Such a model might be useful for developing a qualitative understanding of how mixing different species of DNA functionalized nanoparticles might give rise to networks which are chemically distinct, and in their thermodynamic properties.  Additionally, constraining the number of neighbors in simple spherical potentials is known to dramatically alter the phase behavior~\cite{valency1,valency2,valency3,valency4}. Thus it can be expected that changing the number of DNA strands attached to a core nanoparticle could have a similar dramatic effect. 

\begin{center}
{ \bf ACKNOWLEDGMENTS}
\end{center}

We thank K. Binder, F. Sciortino, Z. Tan, F. Vargas, and D. Wei for
helpful discussions and Wesleyan University for computer time, which was
supported by National Science Foundation Grant CNS-0619508. This work
was supported by National Science Foundation Grant DMR-0427239.

\begin{center}
{\bf APPENDIX}
\end{center}

\begin{figure}
\begin{center}
\includegraphics[width=3.5in]{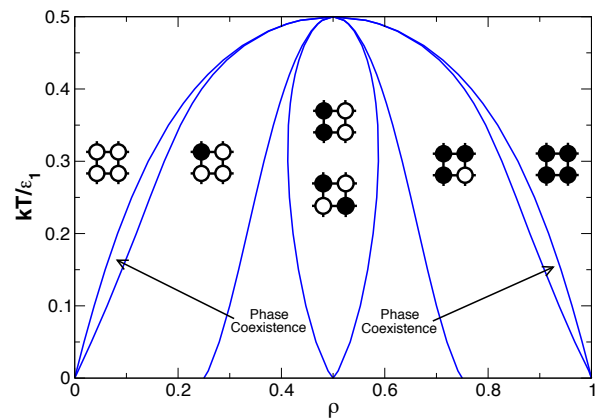}
\end{center}
\caption{Mean-field phase diagram for the case $\epsilon_2/\epsilon_1=1/2$, $\epsilon_1>0$, as a correction for Fig. 6(b) in Ref.~\cite{bl}. All transitions are first order.}
\label{fig:binder}
\end{figure}

During our initial study of lattice gas models with second neighbor
interactions, we attempted to reproduce results of Binder and Landau
for the second neighbor antiferromagnetic case
$\epsilon_2/\epsilon_1=1/2$ and $\epsilon_1<0$. Our results confirmed
the major finding of the earlier work~\cite{bl}. Due to the dramatic
increase in computing power over 28 years since Ref.~\cite{bl} was
published, we could examine the phase behavior much more finely. As a
result, we found that Binder and Landau identified second order
transitions which are in fact actually first order. Specifically, the
highest and lowest density transitions were misidentified in
Ref.~\cite{bl}. We provide a corrected version of this phase diagram in
Fig.~\ref{fig:binder} (compared with Fig. 6(b) in Ref.~\cite{bl}). The
jump in the slope of free energy is too small to be detected with the
accuracy of computation at that time.


\eject




\end{document}